\documentclass[%
aps,
prd,
twocolumn,
preprintnumbers,
nofootinbib,
amsmath,amssymb,
]{revtex4-2}

\usepackage{style}

\begin{document}

\preprint{DESY-25-116}

\title{Self-Gravity in Superradiance Clouds: \\Implications for Binary Dynamics and Observational Prospects}

\author{Hyungjin Kim\,\orcidlink{0000-0002-8843-7690}}
\email{hyungjin.kim@desy.de}
\affiliation{Deutsches Elektronen-Synchrotron DESY, Notkestr. 85, 22607 Hamburg, Germany}

\author{Alessandro Lenoci\,\orcidlink{0000-0002-2209-9262}}
\email{alessandro.lenoci@mail.huji.ac.il}
\affiliation{Racah Institute of Physics, Hebrew University of Jerusalem, 91904 Jerusalem, Israel}
\affiliation{Laboratory for Elementary Particle Physics,
 Cornell University, Ithaca, NY 14853, USA}
 
\begin{abstract}
Spinning black holes could produce ultralight particles via the superradiance instability. 
These particles form a dense cloud around the host black hole, introducing new opportunities for the detection of ultralight new physics. When the black hole is part of a binary system, the binary can trigger transitions among different states of the cloud configuration. Such transitions backreact on the orbital dynamics, modifying the frequency evolution of the emitted gravitational waves. Based on this observation, black hole binaries were proposed as a way to test the existence of ultralight particles. We investigate the effects of the self-gravity of the cloud on the orbital evolution and on the gravitational wave emission. We find that cloud self-gravity could lead to a density-dependent modification of the energy levels of ultralight particles and that it could alter the order of hyperfine energy levels. The crossing of hyperfine levels prevents binaries from triggering resonant hyperfine transitions, and allows them to approach radii that could trigger resonant transitions of fine levels. We study the implications of these findings, especially in the context of future space-borne gravitational wave observatory, the Laser Interferometer Space Antenna (LISA). For quasi-circular, prograde, and equatorial orbits, we find that LISA could probe ultralight particles in the mass range $10^{-15}\eV \, \textrm{--} \, 10^{-13}\eV$ through gravitational wave observations. 
\end{abstract}

\maketitle

\tableofcontents
\smallskip

\section{Introduction}
Ultralight particles appear ubiquitously in numerous beyond the standard model scenarios. A canonical example is the quantum chromodynamics (QCD) axion as a solution to the strong CP problem~\cite{Peccei:1977hh, Peccei:1977ur, Weinberg:1977ma, Wilczek:1977pj, Kim:1979if, Shifman:1979if, Zhitnitsky:1980tq, Dine:1981rt}. Additionally, axion-like particles and dark photons are often considered as benchmark models for phenomenological studies of ultralight new physics. Some of them are associated with theoretical motivations such as solving the electroweak hierarchy problem~\cite{Graham:2015cka, Arvanitaki:2016xds, Geller:2018xvz, Arkani-Hamed:2020yna, TitoDAgnolo:2021nhd, TitoDAgnolo:2021pjo, Chattopadhyay:2024rha}. They might also constitute the dark matter in the present universe~\cite{Banerjee:2018xmn, Banerjee:2020kww, Chatrchyan:2022dpy}. 

Black holes (BHs) provide interesting ways to probe ultralight new physics. 
A spinning black hole can, through the superradiance instability, produce a dense cloud of ultralight particles whose Compton wavelength matches its size.
This process extracts the angular momentum of the black hole, limiting its maximum spin~\cite{Arvanitaki:2009fg}. At the same time, a cloud of ultralight particles could source continuous gravitational waves (GWs) through annihilation and spontaneous emission~\cite{Arvanitaki:2009fg}. This observation leads to a series of surveys to probe ultralight fields via black hole spin measurements and searches for continuous gravitational wave emission generated by these ultralight particles~\cite{Arvanitaki:2010sy, Arvanitaki:2014wva, Arvanitaki:2016qwi, Stott:2018opm, KAGRA:2022osp, Hoof:2024quk, Witte:2024drg, Khalaf:2024nwc, Aswathi:2025nxa, Caputo:2025oap}. 

Another interesting proposal is to use black hole binaries to search for ultralight new physics~\cite{Baumann:2018vus, Baumann:2019ztm}. In the non-relativistic limit, the cloud-BH system is often described as a gravitational atom, analogous to a hydrogen atom, where the black hole serves as the proton and the cloud acts as the electron. When a superradiating black hole forms a binary, the secondary object can trigger a resonant transition between cloud states whose level spacing matches the orbital frequency of the binary. Such transitions then backreact on the orbital dynamics. Depending on the types of transitions and the orientation of orbits, the binary may harden faster or slower than in cases without the superrandiance cloud. This leaves nontrivial time-dependent signatures in the emitted gravitational waves, from which one might infer the existence of ultralight particles. 

\begin{figure}
\centering
\includegraphics[width=.47\textwidth]{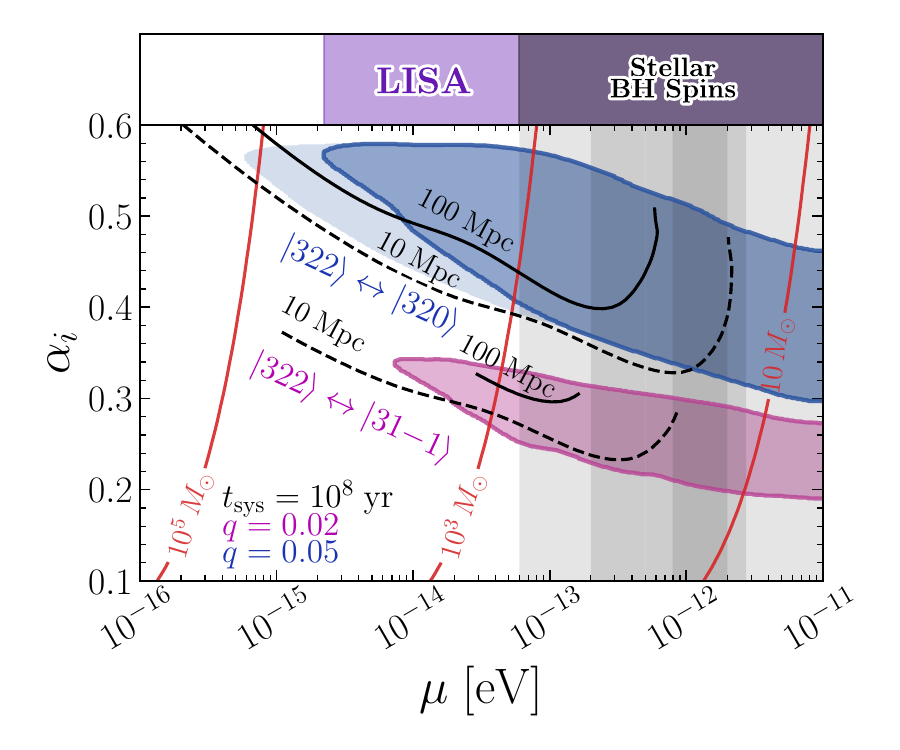}
\caption{Parameter space showing current constraints and the region in which ultralight particles can be probed with LISA for a total observational time span $T_{\rm obs} = 4\ {\rm yr}$. Shaded area indicates regions where LISA is sensitive to ultralight particles via observations of GWs emitted at the hyperfine resonance  $|322\rangle \leftrightarrow |320\rangle$ (blue) and at the fine resonance $|322\rangle \leftrightarrow |31-1\rangle$ (purple). These regions are based on the computation of fitting factor, which will be discussed in Sec.~\ref{sec:observational_target}. The region shaded in lighter blue is where our approximation of neglecting off-diagonal matrix element of the self-gravity breaks down (see Sec.~\ref{sec:off_diagonal}). The fine resonance can be reached due to the level crossing induced by the self-gravity of the cloud. The black contours show the horizon distance at which LISA can observe emitted GWs with  ${\rm SNR}=5$,  while the red contours shows the mass of the spinning black hole $M_1$. We consider only quasi-circular, prograde, equatorial orbits and assume the mass of secondary object $q = M_2/M_1 =0.05$ for the hyperfine transition and $q =0.02$ for the fine transition. Constraints from black hole spin-down are overlaid as vertical gray bands~\cite{Arvanitaki:2014wva, Hoof:2024quk,  Aswathi:2025nxa, Caputo:2025oap}.}
\label{fig:summary}
\end{figure}

In this work, we examine the impact of the self-gravity of the cloud on the orbital dynamics and the emission of gravitational waves. By self-gravity, we refer to the gravitational potential of the cloud itself. We find that self-gravity introduces density-dependent energy level corrections and that this leads to the crossing of certain hyperfine energy levels of the gravitational atom. As a consequence, a binary undergoes a sequence of resonances distinct from the one without self-gravity effects. 

We study the implications of these findings, especially in the context of the future space-borne gravitational wave interferometer LISA. Focusing on quasi-circular, prograde, and equatorial orbits, we find that the self-gravity-induced level crossing allows the binary to enter fine resonances which occur closer to the central rotating black hole. This widens the observational prospects of ultralight particles because gravitational waves emitted by harder binaries are louder and exhibit a faster frequency evolution. In Figure~\ref{fig:summary}, we summarize one of the main findings of this work --- the parameter space, where we could potentially probe the existence of a superradiance cloud through the observation of gravitational waves in LISA. This result is obtained under several requirements, such as the frequency of gravitational wave falling within the LISA frequency band and the waveform being distinguishable from those without a superradiance cloud, among others. The result suggests that ultralight particles could be probed with LISA in the unexplored mass range of $\mu = 10^{-15} \, \textrm{--} \, 10^{-13}\eV$. Details will be presented in the following sections. 

This work is organized as follows. In Section~\ref{sec:review}, we review the basic features of the superradiance cloud and the idea of using binary black holes to probe the existence of ultralight new physics. In Section~\ref{sec:self_gr}, we investigate the impact of self-gravitational effects of the cloud on binary dynamics. In particular, we show that it introduces density-dependent corrections to the energy spectrum and that it could induce crossing among hyperfine levels. In Section~\ref{sec:observational_target}, we discuss the observational implications of these findings, particularly focusing on the future space-borne gravitational wave detector LISA. In Section~\ref{sec:discussion}, we discuss assumptions and simplifications made in the main text that could potentially alter the conclusion of the work. We conclude in Section~\ref{sec:conclude}. Throughout this work, we choose the natural unit $c = \hbar =1$ and the mostly positive metric signature $\eta = (-+++)$.

\section{Review}\label{sec:review}
We review the binary dynamics in the presence of a superradiance cloud. We begin with the basic properties of superradiance instability in Section~\ref{sec:superrad}, and proceed to discuss the idea of using the binary system as a way to probe ultralight new physics~\cite{Baumann:2018vus, Baumann:2019ztm} in Section~\ref{sec:binary}.

\subsection{Superradiance}\label{sec:superrad}
We consider a light scalar particle in the non-relativistic limit. The action for the scalar field is 
\begin{align}
S = \int d^4x \sqrt{-g}
\left[ 
- \frac{1}{2} g^{\mu\nu} \partial_\mu \phi \partial_\nu \phi 
- \frac{1}{2} \mu^2\phi^2 
\right],
\label{action}
\end{align}
where $g_{\mu\nu}$ is the Kerr metric and $\mu$ is the mass of scalar particle. We do not consider the self-interaction in this work. In the non-relativistic limit, the scalar field can be expanded as
\begin{align}
\phi(t,{\bf x} ) = \frac{1}{\sqrt{2\mu}} e^{-i \mu t} \psi(t,{\bf x}) + {\rm h.c.} \, . 
\end{align}
The Klein-Gordon equation for $\phi$ can be written in the form of Schr{\"o}dinger equation,
\begin{align}
i \dot{\psi} 
& \approx
\left( - \frac{\nabla^2}{2 \mu} - \frac{\alpha}{r} \right)\psi 
= H_0 \psi, 
\label{unperturbed}
\end{align}
where $\alpha = G M_1 \mu$ is the fine structure constant of the system and $M_1$ is the black hole mass. The system resembles the hydrogen atom, and for this reason the cloud-BH system is often referred to as a gravitational atom. Here the Kerr metric is expanded to the leading order in $\alpha$; higher order corrections lead to fine and hyperfine splitting of energy levels. 

The spectrum of ultralight particles is similar to that of the hydrogen atom. It consists of a discrete and a continuous spectrum. The discrete spectrum is characterized by three integer quantum numbers; the principal, angular, and magnetic quantum number, $(n,\ell,m)$. The discrete energy spectrum up to ${\cal O}(\alpha^5)$ is given by~\cite{Baumann:2019eav}
\begin{equation}\label{level}
\begin{aligned}
\frac{E_{n\ell m}}{\mu} &= 
1 - \frac{\alpha^2}{2n^2} 
\\
&- \frac{\alpha^4}{8n^4} + \frac{(2\ell -3n +1) \alpha^4}{n^4 (\ell +1/2)} 
\\
&+ \frac{2 a_* m \alpha^5}{n^3 \ell(\ell+1/2)(\ell+1)}\ .
\end{aligned}
\end{equation}
Transitions between two levels, $(n,\ell,m)$ and $(n',\ell',m')$, can be categorized according to the change of quantum numbers:

\medskip\noindent\textbullet~\emph{Bohr transitions:}  transitions between energy levels with different principal quantum numbers, $n\neq n'$. The level spacing is $\Delta E_{\rm Bohr} = {\cal O}(\alpha^2)$. 

\medskip\noindent\textbullet~\emph{Fine transitions:}  transitions between energy levels with the same principal quantum number but with different angular quantum numbers, i.e., $n= n'$ and $\ell \neq \ell'$. The level spacing is $\Delta E_{\rm fine} = {\cal O}(\alpha^4)$. 

\medskip\noindent\textbullet~\emph{Hyperfine transitions:} transitions between levels with the same principal and angular quantum number but with different magnetic quantum numbers, i.e. $n = n'$, $\ell = \ell'$, and $m \neq m'$. The level spacing is $\Delta E_{\rm hyper} = {\cal O}(a_* \alpha^5)$, where $a_* = J/(GM_1^2)$ is the dimensionless spin parameter. 

\medskip 

\noindent This work will focus on fine and hyperfine transitions, in which the self-gravity effect is more relevant. 

\smallskip

The spectrum also develops an imaginary part due to the boundary condition at the black hole horizon. Consequently, the eigenfrequency of the system is
$$
\omega_{n\ell m} = E_{n\ell m} + i\Gamma_{n\ell m},
$$
where the imaginary part $\Gamma_{n\ell m}$ is given by~\cite{Detweiler:1980uk, Baumann:2019eav},
$$
\Gamma_{n\ell m} \propto (m \Omega_+ - \omega_{n\ell m}) . 
$$
Here $\Omega_+ = a_* / [2 r_g ( 1 + \sqrt{1-a_*^2})]$ with $r_g= GM_1$.  When $\Gamma_{n\ell m}>0$, superradiance instability occurs, leading to an exponential production of  ultralight particles. Conversely, when $\Gamma_{n\ell m}<0$, the cloud decays back to the black hole. The superradiance instability occurs only for those states with magnetic quantum number aligned in the direction of the spin axis, i.e., $m>0$, hence the process extracts angular momentum from the black hole. This continues until the black hole spins down enough such that $\Gamma_{n\ell m} \propto (m \Omega_+ -\mu) \approx 0$. 

\begin{figure}
\centering
\includegraphics[width=0.47\textwidth]{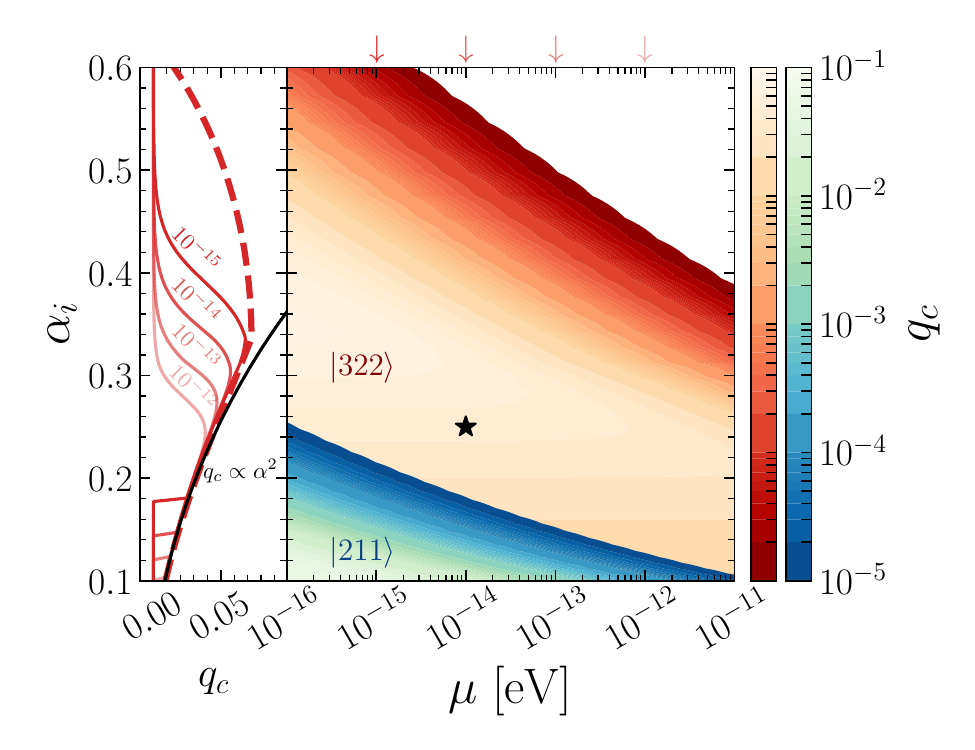}
\caption{Cloud mass fraction $q_c = M_c/M_1$ for $t_{\rm sys} =100\,{\rm Myr}$ and initial  BH spin $a_*^i=0.9$. The right panel shows the cloud mass fraction $q_c$ for $|211\rangle$ and $|322\rangle$ states. We only show the region with $q_c > 10^{-5}$. The upper boundary of the contours is due to the annihilation of the cloud into gravitational waves, while the lower boundary arises because the age of the system $t_{\rm sys}$ is too short for the superradiance instability to develop. The star corresponds to a benchmark point for which the cloud evolution is studied as a function of time in Figure~\ref{fig:qc_max_100Myr_2}. The left panel shows the cloud mass fraction of $|322\rangle$ at $t_{\rm sys} = 100\,$Myr as a function of $\alpha_i$ for $\mu = 10^{-15}\,\textrm{--} \, 10^{-12}\ {\rm eV}$. The red dashed line shows the maximum achievable cloud mass without the annihilation of bosons into gravitational waves. The black line shows the behavior of $q_c^{\rm max}\propto \alpha^2$ for small fine structure constants. See Appendix~\ref{app:cloud_mass} for details on the cloud mass computation. }
\label{fig:qc}
\end{figure}

For a wide range of fine structure constants and black hole spin parameters, the superradiance instability predominantly produces either $|211\rangle$ or $|322\rangle$. We confirm this in Figure~\ref{fig:qc} by computing the ratio between the cloud mass and the black hole mass $q_c = M_c / M_1$ for each state. For the figure, we choose the age of the system $t_{\rm sys} = 100\,{\rm Myr}$, and initial black hole spin $a_*=0.9$. We then numerically solve a set of equations for $M_1$ and $M_c$, which are presented in Appendix~\ref{app:cloud_mass}. The result mildly depends on $t_{\rm sys}$. For a phenomenological reason, we only consider the parameter space where the cloud is dominantly in the $|322\rangle$ state. We will use the above result as an input for the analyses that follow.

\subsection{Binary}\label{sec:binary}
In a binary system, the gravitational atom is tidally perturbed by a secondary object. The Schr{\"o}dinger equation is then given by
\begin{align}
i \dot{\psi}
= \bigg(
- \frac{\nabla^2}{2\mu}
- \frac{\alpha}{r}
+ V_\star
\bigg) \psi , 
\end{align}
where $V_\star$ is the perturbation due to the secondary body,
\begin{align}
V_\star(\boldsymbol r, \boldsymbol r_\star ) 
= - q\alpha
\left[
\frac{1}{| \boldsymbol r - \boldsymbol r_\star (t) |} 
- \frac{1}{r_\star}
- \frac{\boldsymbol r \cdot \boldsymbol r_\star}{r_\star^3}
\right].
\end{align}
Here, $q = M_2 / M_1$ is the ratio between the mass of the rotating black hole and the secondary object, and $\boldsymbol r_\star(t)$ is the position of the secondary object.
The second and third terms in parentheses cancel the monopole and dipole terms in the expansion of the potential around $r=0$. The above Schr{\"o}dinger equation is presented in the black hole comoving coordinate system. A detailed discussion is presented in Appendix~\ref{app:comoving}. 

The system resembles a hydrogen atom with a time-dependent perturbation. The similarity is most clearly illustrated by approximating the system to a two-level system. Consider two levels $\{ | 1\rangle, \, |2\rangle \} = \{ |n_1 \ell_1 m_1\rangle, \, |n_2 \ell_2 m_2\rangle\}$. Each of them is an eigenstate of the unperturbed Hamiltonian, $H_0 |i \rangle = E_i | i \rangle$. We always denote the superradiance state with $|1\rangle$ and a state that can resonate with it via the time-dependent perturbation with $|2\rangle$. A generic state can be written as
$$
|\psi\rangle = c_1(t) | 1 \rangle + c_2(t) | 2\rangle. 
$$
The Schr{\"o}dinger equation is then given by
\begin{align}
i \dot{\boldsymbol c}
= 
\left(
\begin{array}{cc}
E_1 &
\langle 1 | V_\star | 2 \rangle
\\
\langle 2 | V_\star | 1 \rangle 
& E_2
\end{array}
\right) \boldsymbol c , 
\end{align}
where $\boldsymbol c = (c_1 \,\, c_2)^T$. Although the imaginary part of the spectrum is important for the evolution of the system, we ignore it for now for simplicity. 

The time-dependent perturbation $V_\star$ triggers transitions among cloud states. To investigate the transitions, we need to specify an orbit $\boldsymbol r_\star(t)$. We consider a quasi-circular prograde equatorial orbit. The orbit lies in the $xy$-plane, and its angular momentum vector is aligned with the black hole spin direction $\hat z$. We further restrict ourselves to $r_\star > r_B = 1 / \mu\alpha$; the orbit remains outside of the cloud whose size is given by the Bohr radius $r_B$. In this case, the potential can be expanded as
\begin{align}
V_\star(r,r_\star) 
= \sum_{\ell_\star =2} \sum_{m_\star =- \ell_\star}^{\ell_\star} 
V_{\ell_\star m_\star} e^{-i m_\star \phi_\star (t)} , 
\end{align}
where the coefficient $V_{\ell_* m_*}$ is given by
\begin{align}
V_{\ell_\star m_\star} = - q \alpha \,
\frac{4\pi}{2\ell_\star+1} \frac{r^{\ell_\star}}{r_\star^{\ell_\star+1}} Y_{\ell_\star m_\star}(\hat r) Y_{\ell_\star m_\star}^*(\pi/2,0),
\end{align}
and the orbital phase $\phi_\star(t)$ is given by
\begin{align}
\phi_\star(t)
= \pm \int^t dt' \, \Omega(t'), 
\label{phase}
\end{align}
with the orbital frequency $\Omega(t) = \sqrt{GM/r_\star^3(t)}$ and the total mass of the system $M = M_1 + M_2 + M_c$.\footnote{We assume that the center-of-mass of the cloud coincides with the black hole and treat the cloud-BH as a single object constituting the binary system with the secondary object. This will be discussed again in the Appendix.} The plus and minus signs in \eqref{phase} are for prograde and retrograde orbits, respectively. The time-dependence is fully factorized as $\exp[ - i m_\star \phi_\star(t) ]$. The Schr{\"o}dinger equation can be written as
\begin{align}\label{eq:Schrodinger}
i \dot{ \boldsymbol c }
= 
\left(
\begin{array}{cc}
E_1 
& \gamma e^{-i \Delta m_{12} \phi_\star(t) }
\\
\gamma^* e^{+i \Delta m_{12} \phi_\star(t) }
& E_2
\end{array}
\right)
{ \boldsymbol c } , 
\end{align}
where $ \gamma = \sum_{\ell_\star \geq |\Delta m_{12}|} \langle 1 | V_{\ell_\star \Delta m_{12}} | 2 \rangle$ and $ \Delta m_{12} = m_1 - m_2$. This form makes its similarity with a quantum mechanical two-level system transparent. If the orbital frequency is constant, $\phi_\star(t) =  \pm \Omega t$, the resonant transition occurs when $(E_1 - E_2) = \pm \Delta m_{12} \Omega$; the secondary object in the binary plays the role of a laser in resonant transitions of the gravitational atom. In reality, the orbital frequency slowly drifts to a higher value as the binary hardens through gravitational wave emission. Note also that a prograde orbit excites only levels with $(E_1 - E_2) / (m_1 - m_2) >0$, while a retrograde orbit excites levels with $(E_1 - E_2) / (m_1 - m_2) < 0$.

These resonant transitions backreact on the orbital evolution. As the resonant transition changes the angular momentum of the cloud and as the total angular momentum must be conserved, the orbit may decay faster or slower in the presence of the cloud-binary interaction. Whether the orbit decays faster or slower depends on the nature of excitation and the orientation of the orbit. From the angular momentum balance equation (see Appendix~\ref{app:balance}), one finds the orbital frequency evolution equation as
\begin{align}
\frac{d\Omega}{dt} 
=
\left( \frac{d\Omega}{dt} \right)_{\rm GW}
+ \left( \frac{d\Omega}{dt} \right)_{\rm cl} , 
\end{align}
where $(d\Omega/dt)_{\rm GW}$ and $(d\Omega/dt)_{\rm cl}$ each denotes the change of orbital frequency due to the gravitational wave emission and due to cloud internal transitions, respectively. Each of them is given by
\begin{align}
&\left( \frac{d\Omega}{dt} \right)_{\rm GW} 
\!\!\!\!\!= 
+ \frac{96}{5} (G {\cal M}_c)^{5/3} \Omega^{11/3} ,
\label{gw_loss}
\\
&\left( \frac{d\Omega}{dt} \right)_{\rm cl} 
= 
\pm \frac{3 \Omega^{4/3} M_c }{(G {\cal M}_c)^{2/3}} 
\sum_i \frac{m_i}{\mu} 
\left[
\frac{d |c_i|^2}{dt} - 2 \Gamma_i |c_i|^2 
\right] , 
\label{cloud_loss}
\end{align}
where ${\cal M}_c = [ (M_1 +M_c) M_2 ]^{3/5}/ M^{1/5}$ is the chirp mass. The sign in \eqref{cloud_loss} is determined by the relative orientation of the orbital angular momentum with respect to the black hole spin; prograde (retrograde) orbits take the $+$ ($-$) sign. Ignoring the decay rate of the cloud states $\Gamma_i$ and focusing on a two-level system with a prograde orbit, the cloud contribution can be written as
\begin{align}
\left( \frac{d\Omega}{dt} \right)_{\rm cl} 
\propto (m_1 - m_2) \frac{d |c_1|^2}{dt} . 
\end{align}
Since the hyperfine and fine transitions of $|322\rangle$ have $(E_1 - E_2) / (m_1 -m_2)>0$, a resonance can only be triggered by a prograde orbit. Since $(m_1 - m_2)>0$ for both fine and hyperfine transitions, $(d\Omega/dt)_{\rm cl}<0$. The backreaction triggers a floating orbital behavior; the orbit decays slower as the cloud pumps its angular momentum into the binary system~\cite{Baumann:2019ztm}. 

\begin{figure}[t]
\centering
\qquad\includegraphics[width=0.4\textwidth]{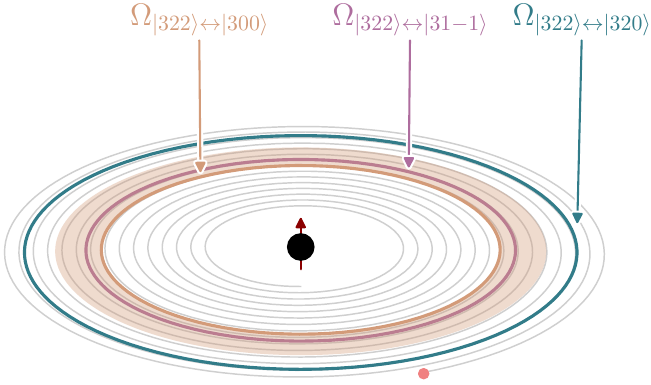}
\\[0.5cm]
\!\!\!\! \includegraphics[width=0.4\textwidth]{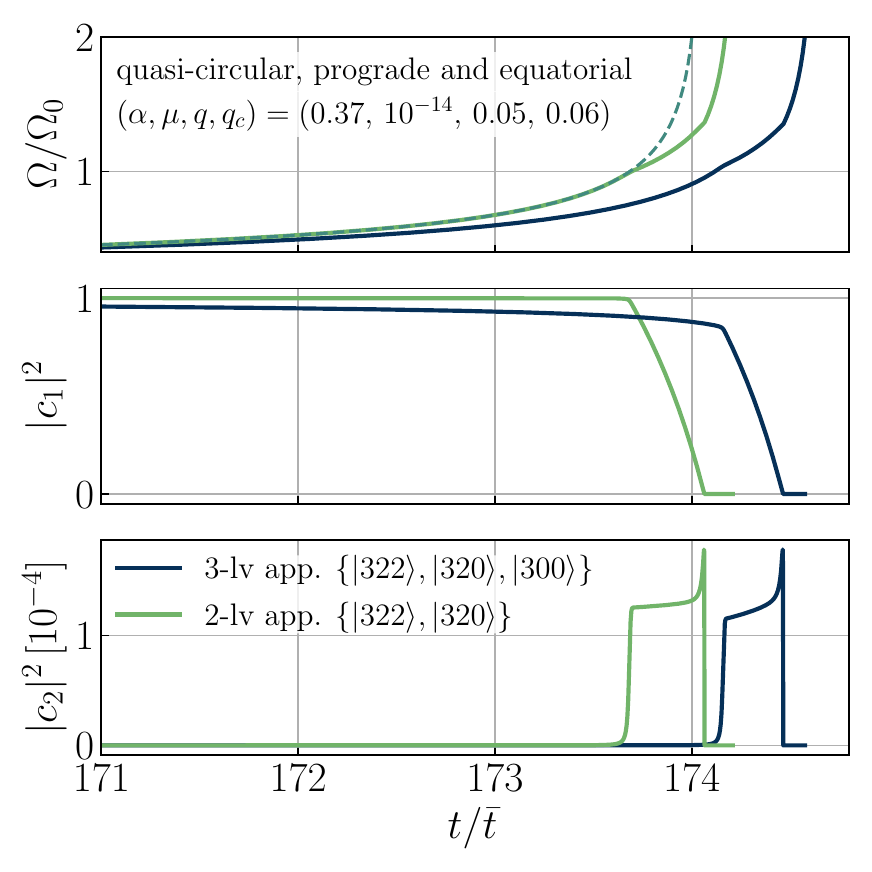}
\caption{(Top) a schematic picture describing the sequence of important orbital resonances with an initial $|322\rangle$ cloud. The orange band around $|322\rangle\leftrightarrow |300\rangle$ denotes the radii around which the mixing with $|300\rangle$ significantly backreacts to the orbit and a three-level description is necessary. This will be discussed in Section~\ref{sec:mixing}. (Bottom) the orbital dynamics at the resonance $|322\rangle \leftrightarrow |320\rangle$ without self-gravity corrections. Compared to the evolution without the cloud (dashed line), the binary hardens much more slowly. The green line denotes the evolution obtained in the two-level approximation, while the dark blue line is obtained in a three-level system, including $|300\rangle$. As this hyperfine transition is adiabatic, the orbit almost completely converts $|322\rangle \to |320\rangle$ and $|320\rangle$ decays subsequently. By the time the orbit reaches to orbital separations that can trigger fine transitions such as the $|322\rangle\leftrightarrow|31-1\rangle$ resonance, the entire cloud has disappeared; this conclusion will be altered when the self-gravity correction is included. Here $\Omega_0$ is the resonance frequency computed with the parameters at the beginning of the evolution and $\bar t = [\Omega(d\Omega/dt)_{\rm GW}^{-1}]|_{\Omega_0}$ is the typical evolution time scale for the orbit due to the GW emission. 
}
\label{fig:floating}
\end{figure}

Figure~\ref{fig:floating} shows a schematic picture of the sequence of orbital resonances considered in this work and an example of a floating orbit. For a quasi-circular prograde orbit, the resonance $|322\rangle \leftrightarrow |320\rangle$ is triggered first among others. This transition tends to deplete the cloud almost entirely, as one can see in the bottom panel of the figure. We will see in the next section that self-gravity alters this behavior. The mixing with the non-superradiance state $|300\rangle$ is important both for hyperfine and fine transitions, which will be discussed in more detail in Section~\ref{sec:mixing}. The floating behavior is slightly different from the ones presented in previous literature. The difference can be attributed to the change of mass and spin of the black hole due to the decay of cloud states; the result in Figure~\ref{fig:floating} is obtained by solving the Schr{\"o}dinger equation, the angular momentum balance equation, and the equation for the black hole and cloud mass simultaneously. Note that $\Omega_0$ is the resonance frequency of the $|322\rangle\leftrightarrow |320\rangle$ transition computed with values of $\alpha$ and $a_*$ at the beginning of the numerical evaluation. 

\section{Self-Gravity}\label{sec:self_gr}
Ultralight particles interact among themselves through the gravitational interaction. This gravitational self-interaction perturbs the Hamiltonian as
\begin{align}
i \dot{\psi} 
= \left(
- \frac{\nabla^2}{2\mu}
- \frac{\alpha}{r}
+ V_c
\right) \psi , 
\label{self_gr_schrodinger}
\end{align}
where the potential due to the self-interaction $V_c(t, \boldsymbol r)$ is given by
\begin{align}
    V_c({t,\boldsymbol r}) 
    = 
    - q_c \alpha
    \int d^3 r' |\psi(t, \boldsymbol r')|^2
    \left[
    \frac{1}{|{\boldsymbol r} - {\boldsymbol r'}|}
    - \frac{1}{r'}
    - \frac{\boldsymbol r \cdot \boldsymbol r'}{r'^3}
    \right]. 
\end{align}
The system is now described by a non-linear Schr{\"o}dinger equation. Note $q_c = M_c / M_1$. The additional terms in parentheses appear due to our coordinate choice.

As the cloud mass could easily be a few percent of the massive black hole, the corrections of the energy spectrum might be large enough so that it becomes comparable to the hyperfine and fine level splitting. Parametrically, the self-gravity correction is $\Delta E_{\rm self} \sim q_c \alpha / r_B$ where $r_B = 1/ \mu \alpha$ is the Bohr radius of the gravitational atom. As already shown in Figure~\ref{fig:qc}, the maximum cloud mass fraction scales as $q_c \propto \alpha^2$~\cite{Baryakhtar:2020gao} (see Appendix~\ref{app:cloud_mass}), and therefore the self-gravity correction could be as large as $\Delta E_{\rm self} \sim q_c \alpha / r_B \sim \mu \alpha^4$. This is of the same order as the fine splitting $\Delta E_{\rm fine} = \mu \alpha^4$, and parametrically larger than the hyperfine splitting $\Delta E_{\rm hyper} = \mu a_* \alpha^5$. The Bohr levels are barely affected.

For a quantitative analysis, we assume an axisymmetric system. In particular, we assume that the cloud is initially in a pure state of $|322\rangle$. With this assumption, we compute the correction to the energy level of each state as
$$
\Delta E_{n\ell m} = \langle n \ell m | V_c | n \ell m \rangle. 
$$
As the cloud has a non-trivial angular distribution, this correction is generally non-universal for states with different quantum numbers. At the same time, the correction depends on the total cloud mass.

\begin{figure}[t]
\centering
\includegraphics[width=0.47\textwidth]{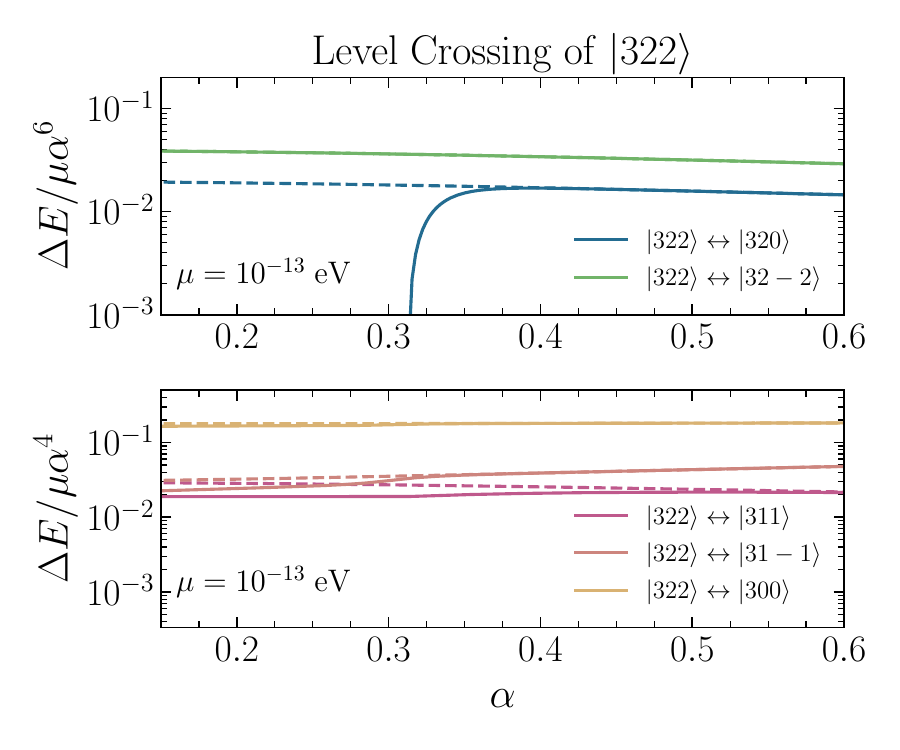}
\caption{
The energy level difference between the $|322\rangle$ state and a few other states. The level crossing occurs for the $|322\rangle \leftrightarrow |320\rangle$ transition for $\mu = 10^{-13}\ {\rm eV}$ around $\alpha \sim 0.31$. Fine transition levels are affected at most at 30\%. We choose the value of $q_c$ with an initial spin parameter $a_* = 0.9$ and $t_{\rm age} =100\,{\rm Myr}$, and use the non-relativistic spectrum for this result. Dashed lines show the level spacing without the self-gravity corrections.}
\label{fig:level_crossing}
\end{figure}

Figure~\ref{fig:level_crossing} shows self-gravity corrections for fine and hyperfine transitions of the superradiance $|322\rangle$ state. The solid (dashed) lines show the level spacing with (without) the self-gravity corrections. For this result, we use the non-relativistic spectrum \eqref{level} and the cloud fraction $q_c$ for $\mu = 10^{-13}\eV$, initial spin parameter $a_* = 0.9$, and the age of the system $t_{\rm sys}=100\,{\rm Myr}$. The corrections to $|322\rangle$ and $|32-2\rangle$ are identical due to the reflection symmetry of the system. Fine levels are affected at most by less than a factor of two at small values of $\alpha$, while the hyperfine splitting $|322\rangle \leftrightarrow |320\rangle$ is significantly affected. In particular, these hyperfine levels change their relative order around $\alpha \simeq 0.3 - 0.4$. 

The above discussion ignores a possible mixing between states induced by self-gravity. In axisymmetric systems, self-gravity triggers mixing between levels that share the same magnetic quantum number, e.g. $|322\rangle$ with $|422\rangle$, $|522\rangle$, $|542\rangle$, and so on. Consider the mixing of $|322\rangle$ with $|422\rangle$. Since $|422\rangle$ has a different principal quantum number, the correction to the energy level is suppressed as $| \langle 422 | V_c |322\rangle|^2 / (E_{322} - E_{422})$, which is negligible compared to $\langle 322 |V_c | 322\rangle$. The same conclusion hold for other states. For instance, $|320\rangle$ could mix with $|300\rangle$. Although it is a mixing between fine levels, one can show that the correction due to the mixing $|\langle 320 | V_c | 300\rangle|^2 / (E_{320} - E_{300})$ is still smaller than $\langle 320 | V_c | 320\rangle$ or $\langle 300 | V_c | 300\rangle$ by three orders of magnitude for a wide range of $\alpha$. The results shown in Figure~\ref{fig:level_crossing} are therefore not significantly affected by the mixing of states.

This level crossing offers new observational opportunities. As the hyperfine transition $|322\rangle \leftrightarrow |320\rangle$ induced by the secondary object tends to be adiabatic, the orbit completely transfers the superradiance state $|322\rangle$ to the non-superradiance state $|320\rangle$, which then decays to the black hole. We have already observed this behavior in Figure~\ref{fig:floating}. The only possibility to probe ultralight particles in this case is therefore by observing the gravitational waves emitted by the binary at this resonance. However, for a moderately small $\alpha$, the resonance occurs too far away from the massive black hole, leading to either too small gravitational wave strain or too slow change of the gravitational wave frequency, both of which hinder the detection of ultralight particles via GW observations. 

\begin{figure*}[t]
\centering
\includegraphics[width=0.45\textwidth]{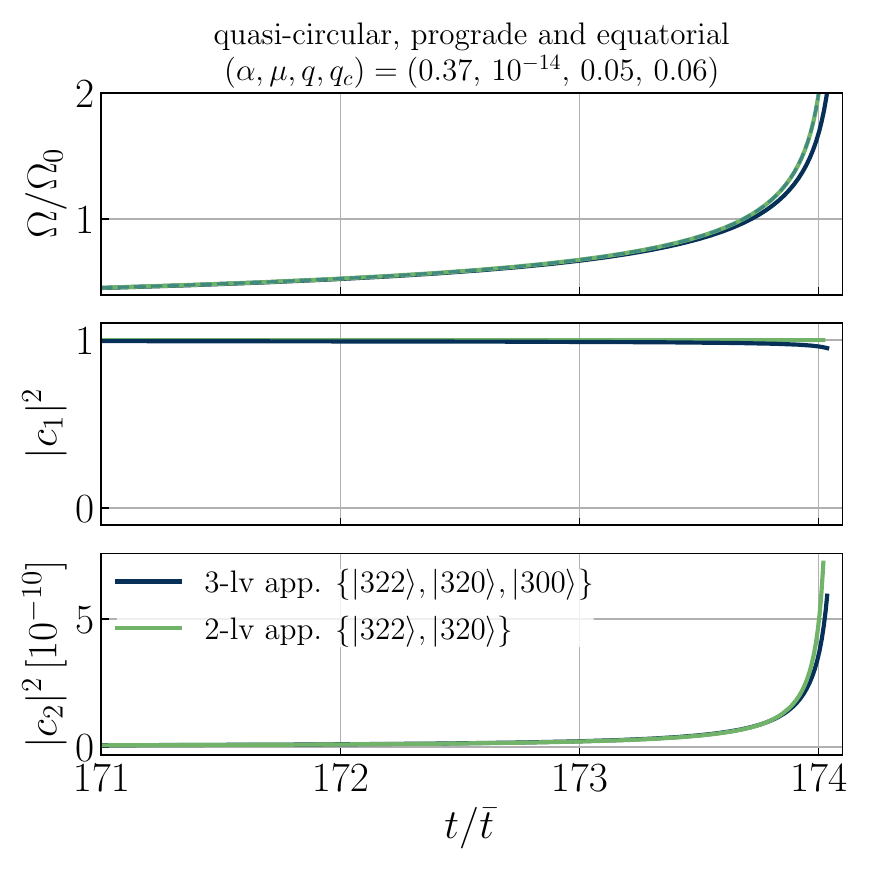}
\qquad
\includegraphics[width=0.45\textwidth]{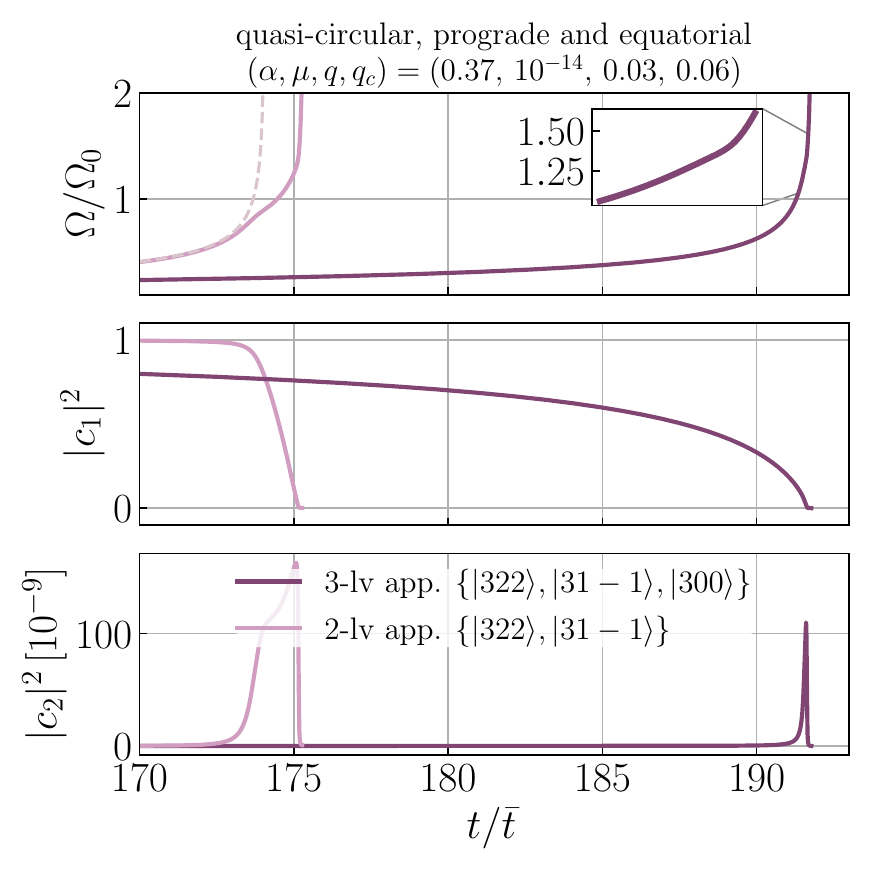}
\caption{(Left) the orbital dynamics at the hyperfine $|322\rangle \leftrightarrow |320\rangle$ resonance. All parameters are chosen the same as in Figure~\ref{fig:floating}. The self-gravity correction is included. As the effective level splitting between these two states changes its sign, a prograde orbit can no longer trigger the resonant transition with them. The cloud still depletes in the 3-level analysis, which is due to the large decay width of the non-superradiance $|300\rangle$ state. The above result suggests that the orbit can reach closer to the black hole, and trigger the resonant transition of the fine levels $|322\rangle \leftrightarrow |31-1\rangle$. (Right) the orbital dynamics around the fine transition $|322\rangle \leftrightarrow |31-1\rangle$. The difference between two-level and three-level approximation is noticeable. The mixing with $|300\rangle$ makes the orbit harden at a much slower rate well before the binary enters the resonance band of the $|322\rangle \leftrightarrow |31-1\rangle$ transition. If the cloud somehow survives by the time it enters the resonance band of the fine transition, there could be another period of evolution in which the binary exhibits a floating behavior as can be seen in the inset plot. For this result, we choose a smaller $q =0.03$. Here $\Omega_0$ is the resonance frequency of each level, computed at the beginning of the numerical evolution with the self-gravity corrections.}
\label{fig:floating_self}
\end{figure*}

With self-gravity correction, the prograde orbit can no longer trigger the resonance transition between $|322\rangle$ and $|320\rangle$ as $[(E_1 + \Delta E_1) - (E_2 + \Delta E_2)] / (m_1 - m_2) <0$ for $\alpha \lesssim 0.3 \, \textrm{--} \, 0.4$. This is shown by the numerical result presented in the left panel of Figure~\ref{fig:floating_self}. The other hyperfine transition $|322\rangle\leftrightarrow|32-2\rangle$ can still resonate with a prograde orbit, but this resonance tends to be non-adiabatic as it can be triggered only with the $\ell=4$ mode of the perturbation $V_\star$, and does not play an important role in the orbital evolution. Not resonating with $|322\rangle\leftrightarrow|320\rangle$, the orbit can approach closer to the black hole, and resonate with fine levels, e.g. $|322\rangle \leftrightarrow |31-1\rangle$. As the fine transitions take place closer to the black hole, the strain and the frequency change of the gravitational wave could potentially be large enough for LISA even with a moderately small $\alpha$, thereby providing another possibility to probe ultralight particles. In the following section, we examine this possibility and investigate in detail the implications of self-gravity for the detection of ultralight bosons with a binary system.

\section{Observational Target}\label{sec:observational_target}
To examine whether we can detect ultralight particles with LISA through the observation of GWs, we focus on two observational targets:
\begin{itemize}
\item GWs emitted at $|322\rangle \leftrightarrow |320\rangle$ with $\alpha \gtrsim 0.3 \, \textrm{--}\, 0.4$ 
\item GWs emitted at $|322\rangle \leftrightarrow |31-1\rangle$ with $\alpha \lesssim 0.3 \, \textrm{--} \, 0.4$ 
\end{itemize}
For the detection of ultralight bosons, two conditions must be met: (i) GWs must be measurable, and (ii) the measurement contains enough information such that it has a discriminating power to distinguish two hypotheses, the one with and the one without ultralight bosons.

The measurability condition (i) is assessed by the following criteria:
\begin{enumerate}
    \item \textbf{GW frequency}.
    The frequency of GWs emitted at the resonance should be within the LISA frequency band, $f_{\rm GW} \in [10^{-5},1]$ Hz.
    
    \item \textbf{GW frequency drift}. 
    The frequency drift of GWs during the observation should be larger than the frequency resolution of the detector, $\Delta f_{\rm GW} > 1/T_{\rm obs}$. Otherwise, the signal is confined to a single frequency bin and can likely be fit without invoking the ultralight cloud, rendering it indistinguishable from GWs emitted by a system without the cloud. For $T_{\rm obs} = 4\,{\rm yr}$, we require $\Delta f_{\rm GW} > 8 \times 10^{-9}$ Hz.
    
    \item \textbf{Signal-to-noise ratio}.
    The signal-to-noise ratio (SNR) should be above threshold for a detection of the signal. We require
    \begin{align}
    \frac{S}{N} = \sqrt{4\int_{f_l}^{f_u} df \, \frac{|h(f)|^2}{S_n(f)}} > \rho_{\rm th} , 
    \label{snr}
    \end{align}
    where $h(f)$ is the GW strain, $S_n(f)$ is the detector noise power spectral density, and $\rho_{\rm th}$ is a pre-defined detection threshold. This condition can be rephrased as a maximum distance to the binary that achieves a detection with ${\rm SNR} = \rho_{\rm th}$. For the computation of the SNR, we use the LISA noise power spectrum in Ref.~\cite{Babak:2021mhe} with the galactic confusion noise presented in Ref.~\cite{Cornish:2017vip}.

\end{enumerate}

The detectability condition (ii) is assessed by the fitting factor
\begin{align}
{\cal F} = {\rm max}_{\boldsymbol \theta_v} 
\frac{
\big( \, 
h(\boldsymbol \theta_{\rm true}) \, \big| \, 
h(\boldsymbol \theta_v) \, 
\big)}{ 
\sqrt{ 
\big( \, 
h(\boldsymbol\theta_{\rm true}) \, \big| \, 
h(\boldsymbol\theta_{\rm true}) \, 
\big) 
\big( \, 
h(\boldsymbol\theta_v) \, \big| \, 
h(\boldsymbol\theta_v) \, 
\big) } }  ,
\label{fitting factor}
\end{align}
where the inner product is defined as
\begin{align}
( A | B ) = 
2 \, {\rm Re} 
\int_{-\infty}^{\infty} df \, \frac{A^*(f) B(f)}{S_n(f)}. 
\label{inner_product}
\end{align}
Here $h(\boldsymbol\theta)$ denotes the waveform of gravitational waves. The waveform is characterized by a set of parameters $\boldsymbol \theta = \{ f_{-}, \, {\cal M}_c, \, q_c, \, \mu, \, \alpha\}$, where $f_-$ is the frequency of GWs at the beginning of the observational campaign. This set only includes the intrinsic variables; the extrinsic variables are already algebraically maximized in the fitting factor. The inner product in \eqref{fitting factor} is maximized only over a subset of parameters $\boldsymbol \theta_v = \{ f_- , {\cal M}_c \}$, assuming that $q_c \to 0$, $\alpha \to \infty$, and $\mu \to \infty$. The resulting $h(\boldsymbol \theta_v)$ represents a waveform of GWs from the system without the cloud. The fitting factor therefore measures how well the vacuum waveform $h(\boldsymbol \theta_v)$ could fit the GW signals emitted from a system with a cloud of ultralight particles.

\begin{figure}
\centering
\includegraphics[width=0.47\textwidth]{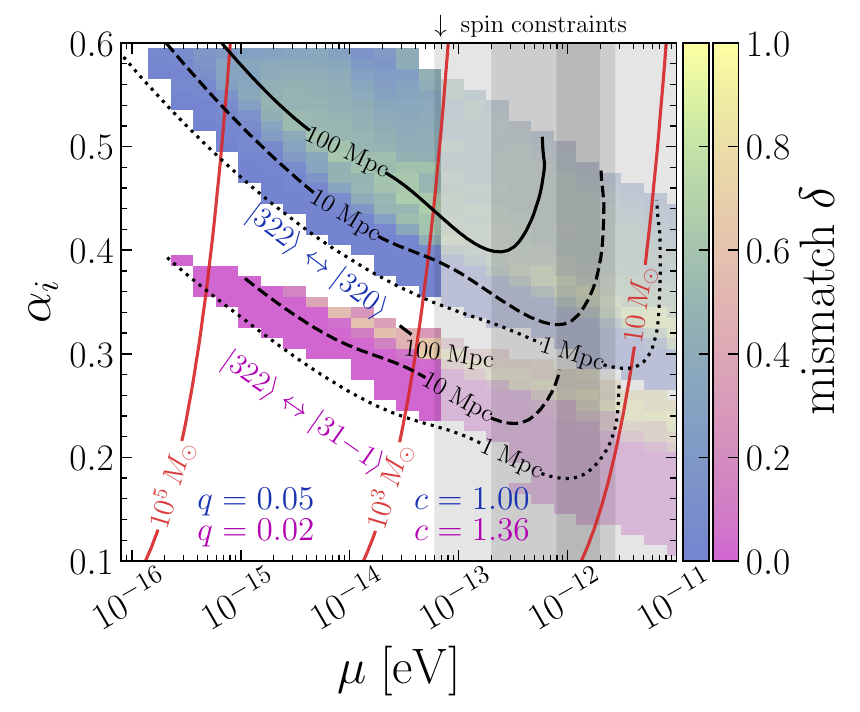}
\caption{Summary of the measurability conditions and the mismatch. The black contours show the maximum distance from which the emitted gravitational waves can be detected at LISA with a signal-to-noise ratio $\rho_{\rm th}=5$. The mismatch $\delta = 1-F$ is overlaid. The purple map shows the mismatch for GWs emitted at the fine resonance $|322\rangle \leftrightarrow |31-1\rangle$, while the blue map shows it for GWs emitted at the hyperfine resonance $|322\rangle \leftrightarrow |320\rangle$. We stress that the fine transition can only be reached for $\alpha \lesssim 0.3 - 0.4$ as a result of the level crossing. We choose $q=0.05$ for the hyperfine and $q=0.02$ for the fine transition. We compute the fitting factor only for the parameter space with $q_c(100\,{\rm Myr})>10^{-6}$. The red contours show the mass of the rotating black hole, while the gray shaded region shows the constraints from black hole spin measurements
~\cite{Arvanitaki:2014wva, Hoof:2024quk,  Aswathi:2025nxa, Caputo:2025oap}.}
\label{fig:322_conditions}
\end{figure}

Figure~\ref{fig:322_conditions} shows the intersection of the three measurability requirements and the fitting factor. The measurability condition is illustrated by the black contours, representing the maximum distance to a source from which gravitational waves can be detected by LISA with a signal-to-noise ratio of $\rho_{\rm th}=5$. For both fine and hyperfine transitions, LISA can measure GWs emitted at the resonances from sources a few tens to hundreds megaparsec away from us. At the same time, we show the fitting factor computed at each point of the parameter space where the measurability conditions are satisfied. In the figure, we show the mismatch
$$
\delta  = 1 - {\cal F} \in [0,1]. 
$$
When two waveforms are orthogonal to each other, the mismatch is $\delta =1$. In this case, the GWs emitted from the binary with and without the cloud can be distinguished from each other. For this figure, we choose $T_{\rm obs}= 4\,{\rm yr}$, and use the numerically obtained cloud mass fraction $q_c(100\,{\rm Myr})$. We assume a mass ratio of the companion and central black hole of $q=0.05$ for the hyperfine transition and of $q=0.02$ for the fine transition. 

To quantify the detectability of the cloud, we require the mismatch to satisfy
\cite{Flanagan:1997kp, Lindblom:2008cm, McWilliams:2010eq, Chatziioannou:2017tdw, Purrer:2019jcp},
\begin{align}
    \delta = 1-{\cal F} > \frac{D}{2 {\rm SNR}^2} ,
    \label{detectability_crit}
\end{align}
where $D$ is the number of parameters fitted in the fitting factor. In our case $D=2$. The right-hand side of this criterion is the mismatch that could arise as a result of pure statistical fluctuations. We therefore impose that the mismatch arising from  genuine differences between two waveforms exceeds the statistical fluctuation. In practice, we impose a stricter condition; the mismatch to be 2$\sigma$ away from null for a conservative estimate. For SNR$\,=5$, we require $\delta = 1 - {\cal F}>0.16$ for the detectability of the cloud.

When computing the fitting factor, we must specify the frequency of the gravitational wave that enters the detector during the observation. Suppose that the observational campaign runs for $t \in [t_0 - 2\,{\rm yr}, t_0 + 2\,{\rm yr}]$, where $t_0$ is the midpoint of the observation. The fitting factor shown in Figure~\ref{fig:322_conditions} is computed by assuming $f_{\rm GW}(t_0) = \Omega_0/\pi$ for the hyperfine resonance and $f_{\rm GW}(t_0) \simeq 1.36 (\Omega_0/\pi)$ for the fine resonance. In other words, we assume that the GWs emitted near each resonance enter at the midpoint of the campaign. As such GWs contain the richest information about the cloud, they provide the clearest way to test the existence of ultralight particles. 

\begin{figure*}[t]
\centering
\includegraphics[width=0.45\textwidth]{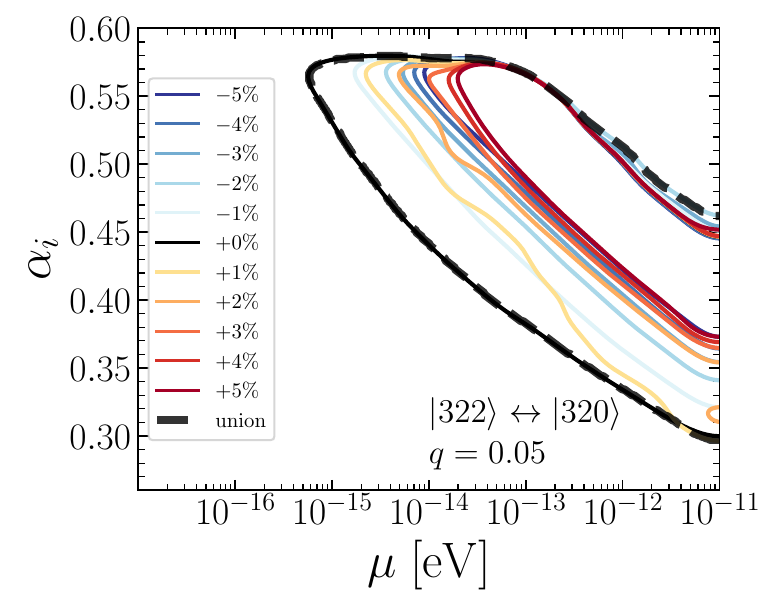}
\includegraphics[width=0.45\textwidth]{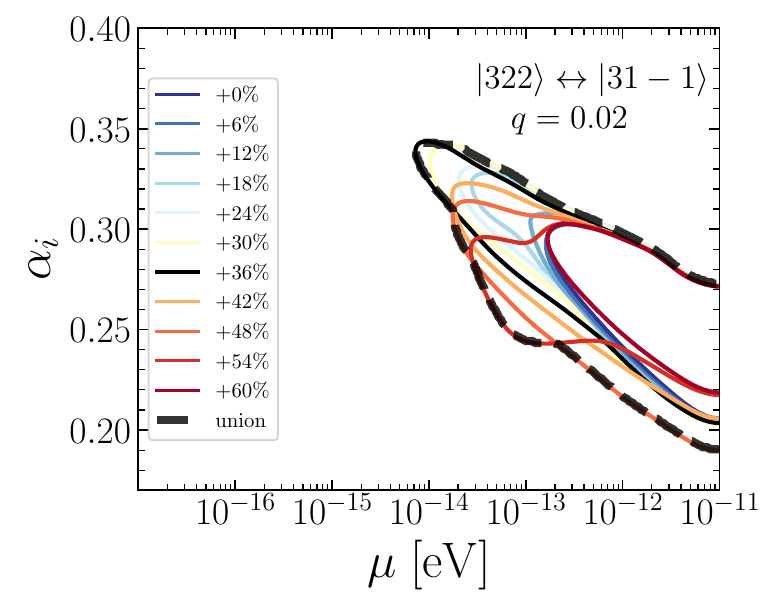}
\caption{Contours of mismatch satisfying $\delta \geq 4 D / 2 {\rm SNR}^2$ for different choices of $c$, defined via $f(t_0) = c (\Omega_0/\pi)$. 
(Left) the mismatch for the hyperfine transition $|322\rangle \leftrightarrow |320\rangle$ with  $q=M_2/M_1 = 0.05$.
The mismatch sharply drops as $c$ drifts away from $c=1$, suggesting that the ability to probe the existence of the cloud crucially depends on whether the gravitational waves emitted at the resonance enter the detector during the observational campaign.
(Right) the mismatch for for the fine transition $|322\rangle \leftrightarrow |31-1\rangle$ with  $q=M_2/M_1 = 0.02$. The results for other combinations of $(c, \, q)$ are presented in Appendix~\ref{app:fitting factor}.}
\label{fig:misaligned_freq}
\end{figure*}

The resulting mismatch depends on the frequency of gravitational waves $f_{\rm GW}(t_0)$ that enters the detector during the observation. To examine how the mismatch changes as a function of $f_{\rm GW}(t_0)$, we repeat the computation with
\begin{align}
f_{\rm GW}(t_0) = c \frac{\Omega_0}{\pi} , 
\end{align}
where $c$ parametrizes the deviation of the GW frequency from the resonance frequency. In Figure~\ref{fig:misaligned_freq}, we show the region of mismatch that satisfies the criterion \eqref{detectability_crit} at $2\sigma$ for different choices of $c$. For the hyperfine transition, the largest region is achieved around $c=1$, confirming that GWs from the resonance contain the most information on the cloud around the rotating black hole. On the other hand, for the fine transition, the largest region is achieved for $c\simeq 1.36$. The reason for this is due to $|300\rangle$; before the binary enters the fine resonance, the cloud depletes due to its mixing with $|300\rangle$, and the resonance frequency increases compared to $\Omega_0$, which is specified at the beginning of the numerical evaluation.  Figure~\ref{fig:summary} is obtained based on Figure~\ref{fig:322_conditions} -- \ref{fig:misaligned_freq}; in particular, the projections in Figure~\ref{fig:summary} are obtained from the union of mismatch contours shown in Figure~\ref{fig:misaligned_freq}. Details on the computation of fitting factor and the results for other values of the mass ratio $q$ are presented in Appendix~\ref{app:fitting factor}. 

Before ending this section, we add a brief comment on the mismatch. The mismatch can be interpreted as a reduction of the signal-to-noise ratio due to a mismodeling of the waveform. To see this, consider a data stream in the strong-signal limit,
$$
s(t) = h_{c}(t; \boldsymbol\theta_{c}) + n(t)
\approx h_c(t; \boldsymbol \theta_{c}) , 
$$
where $n(t)$ is some Gaussian noise, and $h_c(t; \boldsymbol\theta_c)$ is the true gravity wave strain parameterized by $\boldsymbol \theta_c$. The optimal statistic with a template $h(\boldsymbol \theta)$ can be defined as
\begin{align}
\hat \rho 
&= 
\frac{( \, s \, | \, h(\boldsymbol \theta) \, )}{ \sqrt{ ( \, h(\boldsymbol \theta) \, | \, h(\boldsymbol \theta) \, ) } }  .
\end{align}
When the template matches with the true waveform, the maximum signal-to-noise ratio is given by
\begin{align}
\Big( \frac{S}{N} \Big)_{\rm max}
= \max_{\boldsymbol\theta} \langle \hat \rho \rangle
= \big( \, h(\boldsymbol \theta_{\rm true}) \, \big| \, h(\boldsymbol \theta_{\rm true}) \, \big)^{1/2} . 
\end{align}
When one chooses $h(\boldsymbol \theta) = h(\boldsymbol \theta_v)$ which does not match the signal exactly, the maximum signal-to-noise ratio is reduced by
\begin{align}
\frac{S}{N} 
=\max_{\boldsymbol \theta_v} \langle \hat \rho \rangle
= \Big( \frac{S}{N} \Big)_{\rm max} {\cal F}. 
\end{align}
The fitting factor ${\cal F}$ can therefore be understood as the reduction of the maximum signal-to-noise ratio due to mismodelling of the waveform~\cite{Ajith:2007kx}.

\section{Discussion}\label{sec:discussion}

\subsection{Mixing with $3s$}\label{sec:mixing}
The mixing with spectator states (that do not participate to the resonance) could be important for the evolution of the system. A primary example is the mixing of $|322\rangle$ with $|300\rangle$. Due to its quantum numbers, the $3s$ state has a large decay rate $\Gamma_{300} \propto \mu \alpha^5$, and at the same time, it can mix non-resonantly with the state $|322\rangle$ via the quadrupole component of the perturbation $V_\star$. Its importance for orbital dynamics is already hinted in the numerical results presented in previous sections, e.g. Figure~\ref{fig:floating} and \ref{fig:floating_self}. This behavior was already observed in the work of Tong et al~\cite{Tong:2022bbl}. See also Refs.~\cite{Takahashi:2023flk, Tomaselli:2025jfo} for discussions on finite decay width of states and its implication for orbital dynamics.

The mixing introduces a steady decay of the $|322\rangle$ state into the black hole. Consider a two-level system, consisting of $\{ | 1 \rangle, | 3 \rangle\} = \{ | 322 \rangle, | 300 \rangle \}$. With a diagonal phase rotation, one can show that the two-level Hamiltonian~\eqref{eq:Schrodinger} can be written as
\begin{align}
i \dot{\boldsymbol c}
= \left(
\begin{array}{cc}
E_1 - \Delta m_{13} \Omega /2 
& \gamma
\\
\gamma
& E_3 + \Delta m_{13} \Omega/2 + i \Gamma
\end{array}
\right)
{\boldsymbol c},
\end{align}
where $\Omega = \dot\phi_\star(t)$ is the orbital frequency for the prograde orbit, $\Delta m_{13} = m_1 - m_3$,  $\Delta E_{13} = E_1 - E_3$, and $\Gamma$ is the decay rate of the spectator state $|3\rangle$. From this, the mixing angle may be estimated as
\begin{align}
\theta \simeq \frac{\gamma}{ \Delta E_{13} - \Delta m_{13} \Omega - i \Gamma }. 
\end{align}
Even though the state initially begins with the non-decaying $|1\rangle$ state, it finds itself in the decaying $|3\rangle$ state with a probability of $|\theta|^2$, causing a steady decay of the $|1\rangle$ state throughout the evolution, 
\begin{align}
|c_1|^2 \propto \exp\left[ - 2 \int^t dt'\, |\theta|^2 \Gamma\right] . 
\label{induced_decay}
\end{align}
 A more detailed derivation this expression will be given in Appendix~\ref{app:decay}. 

This steady decay is already observed in Figure~\ref{fig:floating}. The occupation number of the superradiance state steadily decreases in the three-level approximation until it reaches the resonance of $|322\rangle \leftrightarrow |320\rangle$. The decay induced by mixing \eqref{induced_decay} reasonably agrees with the numerical result when an additional numerical factor of $1.7$ is introduced in the exponent. This steady decay of the superradiance state could limit the possibility of observing the ultralight cloud through GWs, as it might exhaust the entire cloud well before it reaches resonances~\cite{Tong:2022bbl, Fan:2023jjj}.

Furthermore, the mixing-induced decay backreacts to the orbit. The orbital evolution of a three-level system is retarded compared to that of a two-level system, as shown in Figure~\ref{fig:floating}. This can be explained as follows. The superradiance state dumps its energy into the black hole, $|322\rangle \to |300\rangle \to |{\rm BH}\rangle$, through the mixing enabled by the secondary object and the $|300\rangle$ decay. During this process, the state transfers its angular momentum to the secondary object's orbit, causing the orbit in the three-level system to float longer compared to the one in the two-level system. 

This mixing-induced backreaction causes a more dramatic orbital behavior near the resonance $|322\rangle \leftrightarrow |31-1\rangle$. The right panel of Figure~\ref{fig:floating_self} shows that the orbit floats for a much longer period of time due to its mixing with $|300\rangle$. Contrary to the hyperfine case, the orbit gains a large angular momentum from the mixing-induced cloud evolution as much as it loses via gravitational wave emission. The system then enters a quasi-equilibrium state where the angular momentum gain from the cloud $(d\Omega/dt)_{\rm cl}$ balances the angular momentum loss due to the GW emission $(d\Omega/dt)_{\rm GW}$. This behavior continues until the cloud is fully exhausted. In the case of the right panel of Figure~\ref{fig:floating_self}, a small $|322\rangle$ population survives, enough for the $|322\rangle \leftrightarrow |31-1\rangle$ resonance to be triggered. This is shown as an inset plot in the top panel.

\subsection{Relativistic Corrections}

Throughout this work, we use the non-relativistic approximation, expected to be valid for $\alpha / \ell < 1$. The results in the previous section are shown up to $\alpha = 0.6$, which is still smaller than unity, but not sufficiently. It is therefore important to check if the conclusions of the previous sections still hold at least qualitatively when relativistic corrections are included. 

Relativistic effects could modify the spectrum and wave functions. The change in the real part of the spectrum could affect the sequence of orbital resonances, including the level crossing of $|322\rangle \leftrightarrow |320\rangle$, while the imaginary part of the spectrum could affect the mixing-induced evolution of the BH-cloud system. At the same time, the change in the wave function could modify the matrix element of perturbations, and thereby, affecting the resonant transitions. To quantify these effects, we compute the fully relativistic spectrum and wave function, following the procedure outlined by Dolan~\cite{Dolan:2007mj} with a saturated spin parameter $a_* = 2\alpha / (1 + \alpha^2)$ for the $|322\rangle$ cloud (see Appendix~\ref{app:relativistic} for more details). For illustration, we consider a benchmark with $\mu=10^{-13}$ eV.

\begin{figure}[t]
\centering
\includegraphics[width=0.45\textwidth]{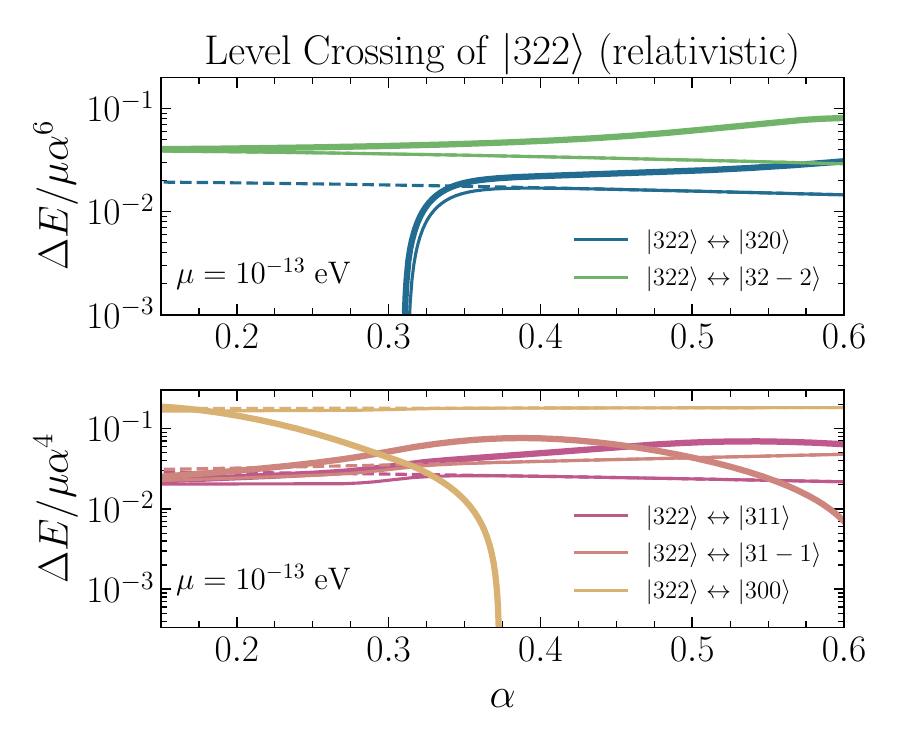}
\!\!\!\!\includegraphics[width=0.45\textwidth]{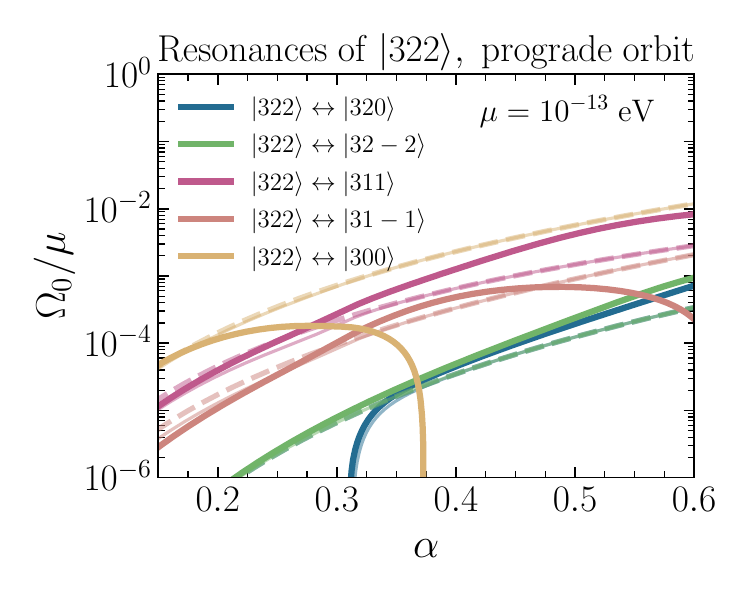}
\caption{
(Top) same as Figure~\ref{fig:level_crossing} but including relativistic corrections. (Bottom) resonant frequencies for the indicated transitions. Thick lines include both self‑gravity and relativistic corrections. Solid opaque lines show the non‑relativistic limit with self‑gravity, while dashed lines show the same limit without self‑gravity.
}
\label{fig:crossing_relativistic}
\end{figure}

The level crossing of $|322\rangle\leftrightarrow |320\rangle$  remains the same. This phenomenon occurs at a relatively small $\alpha < 0.3 \, \textrm{--} \, 0.4$, and thus is expected to be less affected by relativistic effects. This is confirmed by the numerically computed relativistic spectrum shown in the top panel of Figure~\ref{fig:crossing_relativistic}; the level crossing still occurs around $\alpha=0.3$ even after including relativistic corrections to the spectrum.

The orbital resonance sequence remains mostly the same. For fine transitions at small $\alpha$, e.g. $\alpha < 0.3$, the orbit will still trigger the transition $|322\rangle \leftrightarrow |31-1\rangle$ first among the other fine transitions. This can be checked in the bottom panel of Figure~\ref{fig:crossing_relativistic}. For hyperfine transitions, a quasi-circular orbit will trigger $|322\rangle \leftrightarrow |320\rangle$ first among all the others, except for a fine structure constant around $\alpha =0.6$ in which $|322\rangle\leftrightarrow |31-1\rangle$ might be triggered first. 

The evolution due to the mixing with the spectator $|300\rangle$ might be affected by relativistic corrections. First, the decay rate of $|300\rangle$ changes by a factor of few compared to the non-relativistic approximation. Second, the level splitting $\Delta E_{13} = E_{322} - E_{300}$ differs from its non-relativistic counterpart by a factor of few and also changes its sign at $\alpha \simeq 0.37$. Third, the matrix element $\gamma_{13}$ might change due to the relativistic correction to the wave function. Combined together, they could modify the exponent of \eqref{induced_decay}.

\begin{figure}[t]
\centering
\includegraphics[width=0.45\textwidth]{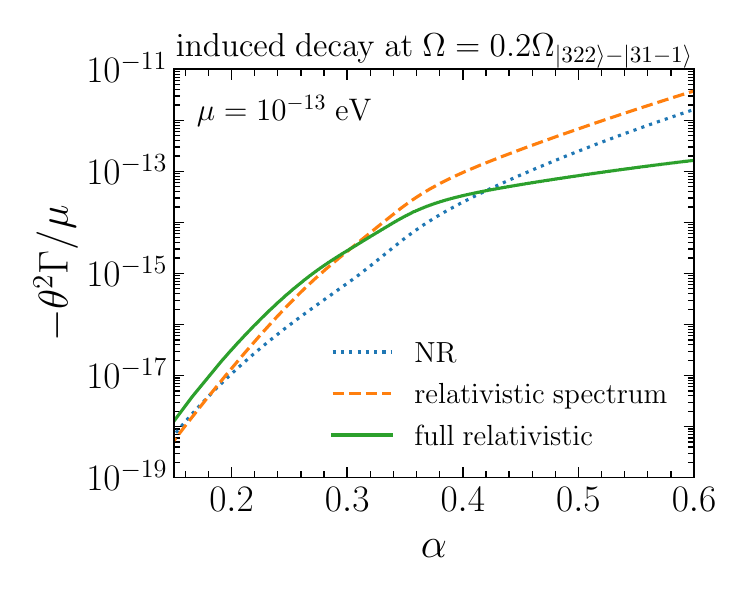}
\caption{The squared mixing angle multiplied with the decay width. We show the non-relativistic result (blue dotted), the relativistic result (green), and the result with relativistic correction to the spectrum but with non-relativistically computed matrix element (orange dashed). We choose $\mu = 10^{-13}$ eV, $q=0.05$, and $\Omega = 0.2 \Omega_{|322\rangle \textrm{--} |31-1\rangle}$ as a reference frequency. Self-gravity corrections to the spectrum are also included.}
\label{fig:induced}
\end{figure}

To examine this, we compute $\theta^2 \Gamma$ near the fine resonance, including relativistic corrections to the decay rate of $|300\rangle$, to the level splitting $E_{322} - E_{300}$, and the matrix element. The result is shown in Figure~\ref{fig:induced}. The relativistic result (green solid) is larger than the non-relativistic one (blue dotted) by a factor of few for $\alpha \lesssim 0.4$. This is mainly due to relativistic corrections to the energy spectrum: the orange dashed line --- computed with relativistic corrections to the decay width and energy spectrum but using the non‑relativistic matrix element $\gamma$ --- almost reproduces the fully relativistic result. At $\alpha\gtrsim0.4$ the relativistic result is up to one order of magnitude smaller than the non-relativistic result due to relativistic corrections of the matrix element (see Figure~\ref{fig:relativistic_matrix} in the Appendix).

The above arguments show that the level crossing behavior as well as the orbital dynamics remains qualitatively the same even after including relativistic corrections. The relativistic corrections considered in this section are however restricted. We only consider relativistic corrections to the energy spectrum and the wave function due to an isolated black hole. There are other corrections we do not include: perturbations nonlinear in the external matter distribution, ${\cal O}(q^n)$ ($n\geq 2$), and perturbations of order ${\cal O}(q \alpha^n)$ ($n > 2$). Here $q$ stands for both $q$ and $q_c$. In the following section, we will consider one of the ignored ones: a perturbation of ${\cal O}(q_c \alpha^4)$ order. Another possible source of error can come from a partial metric reconstruction when the secondary approaches the primary in the presence of the cloud: for the largest values of $\alpha$ considered in this work, the used Kerr metric perturbed with the secondary (see Appendix~\ref{app:tidal_pert} for details) may not be the proper metric to describe the binary system in presence of the cloud. To fully account for all these corrections, a dedicated numerical simulation is required. 

\subsection{Corrections from Cloud Angular Momentum}
The gravitational self-interaction introduces additional corrections to the energy spectrum. Takahashi et al~\cite{Takahashi:2021yhy} showed that the angular momentum of the cloud affects the spectrum at the order of the hyperfine splitting. As it is also due to the gravitational self-interaction, it depends on the occupation number of the cloud itself. In our analysis, we have ignored this correction. We assess below the relative importance of the self-gravity with other self-gravitational corrections. 

We begin with the non-relativistic expansion of the scalar action. With the Kerr metric, the scalar action \eqref{action} can be expanded as
\begin{align}
S = \int d^4x \, 
\big(
{\cal L}_2
+ {\cal L}_4
+ {\cal L}_5
+ \cdots
\big) ,
\label{expansion}
\end{align}
where ${\cal L}_2$ is the leading order Lagrangian that gives rise to the unperturbed Schr{\"o}dinger equation \eqref{unperturbed} and ${\cal L}_4$ is the ${\cal O}(\alpha^4)$ correction corresponding to the fine splitting.
The hyperfine splitting arises from ${\cal L}_5$ term~\cite{Baumann:2018vus}
\begin{align}
{\cal L}_5 = - i g^{t\phi} \psi^* \partial_\phi \psi , 
\end{align}
where $g^{t\phi} \approx - 2 a_* (GM_1)^2 /r^3$. The correction to the time-space component of the metric from the cloud is obtained assuming a flat background~\cite{Takahashi:2021yhy}
\begin{align}
\delta g_{ti}
&= 4 G \int d^3 \boldsymbol x' 
\frac{Q_{ti}(\boldsymbol x')}{|\boldsymbol x - \boldsymbol x'|},
\label{gti}
\end{align}
where $Q_{ti} = \frac{i}{2} (\psi_{\rm cl}^* \partial_i \psi_{\rm cl} - \psi_{\rm cl} \partial_i \psi_{\rm cl}^*)$ and $\psi_{\rm cl}$ is the cloud wave function. The Hamiltonian is corrected by $\Delta H = \delta g^{t\phi} (i \partial_\phi) = - \delta g^{t\phi} L_z$, and the energy correction due to the angular momentum of the cloud is given by
\begin{align}
\Delta E_{n\ell m}
=& - m \langle n \ell m | \delta g^{t\phi} | n\ell m \rangle . 
\end{align}
For the $|322\rangle$ cloud, the level splitting between $|322\rangle$ and $|320\rangle$ due to the cloud angular momentum is given by
\begin{align}
\Delta E_{\rm ang}
= \frac{209}{30240} \mu q_c \alpha^4. 
\end{align}
As it is already expected, this correction is order of ${\cal O}(\alpha^6)$, while the self-gravity correction studied in this work is ${\cal O}(\alpha^4)$. With the explicit result for the self-gravity correction for the $|322\rangle$ and $|320\rangle$ states, $\Delta E_{\rm self} = - (6851/967680) \mu q_c \alpha^2$, which arises from the correction $\delta g_{tt}$, we find
\begin{align}
\frac{\Delta E_{\rm ang}}{\Delta E_{\rm self}}
\simeq - \alpha^2. 
\label{est}
\end{align}
The angular momentum correction remains at most at the level of $\sim 30\%$ for the entire range of fine structure constant, where the hyperfine transition is phenomenologically relevant.\footnote{This estimate should be considered as an approximate one. The correction to the metric $\delta g_{ti}$, when expressed in the black hole comoving coordinate, has no monopole and dipole contributions, and hence \eqref{gti} should be subtracted with $1/x' + \boldsymbol x \cdot \boldsymbol x'/ x'^3$ in the same manner as in the case of $V_\star$ and $V_c$. This reduces the estimate \eqref{est} by two orders of magnitude. At the same time, $\delta g_{ti}$ contains other terms that arise from the coordinate transformation between the barycenter frame and the comoving frame, whose effect is expected to be at the same order.}

\subsection{Off-Diagonal Self-Gravity Matrix Element}\label{sec:off_diagonal}
We have neglected the off-diagonal self-gravity matrix elements $\langle 1 | V_c | 2 \rangle  $ in the numerical evolution of the system. In the following, we offer justifications for this simplified treatment. 

Consider the hyperfine splitting between $|322\rangle$ and $|320\rangle$. We assume that the initial cloud configuration is $|\psi\rangle \propto |322\rangle$. Before the secondary object is introduced, the Hamiltonian of the system is diagonal due to the axisymmetry of the system, i.e. $\langle 322 | V_c | 320 \rangle = 0$. Only when the secondary object is introduced the axisymmetry of the system is explicitly broken and $|322\rangle$ begins to mix possibly with $|320\rangle$. Hence, we expect the size of the off-diagonal element due to the self-gravity to be parametrically suppressed by the relative occupation number of $|320\rangle$ to $|322\rangle$ state. For $\alpha < 0.3 \, \textrm{--} \, 0.4$, resonant mixing is not possible in the first place due to the crossing of levels, and therefore we expect that the off-diagonal element $\langle 322 | V_c |320\rangle$ is irrelevant for the dynamical evolution of the system across the hyperfine splitting. For $\alpha > 0.3 \, \textrm{--} \, 0.4$, resonant mixing is possible, which could lead to a non-vanishing off-diagonal matrix element. 

\begin{figure}
    \centering
\includegraphics[width=0.47\textwidth]{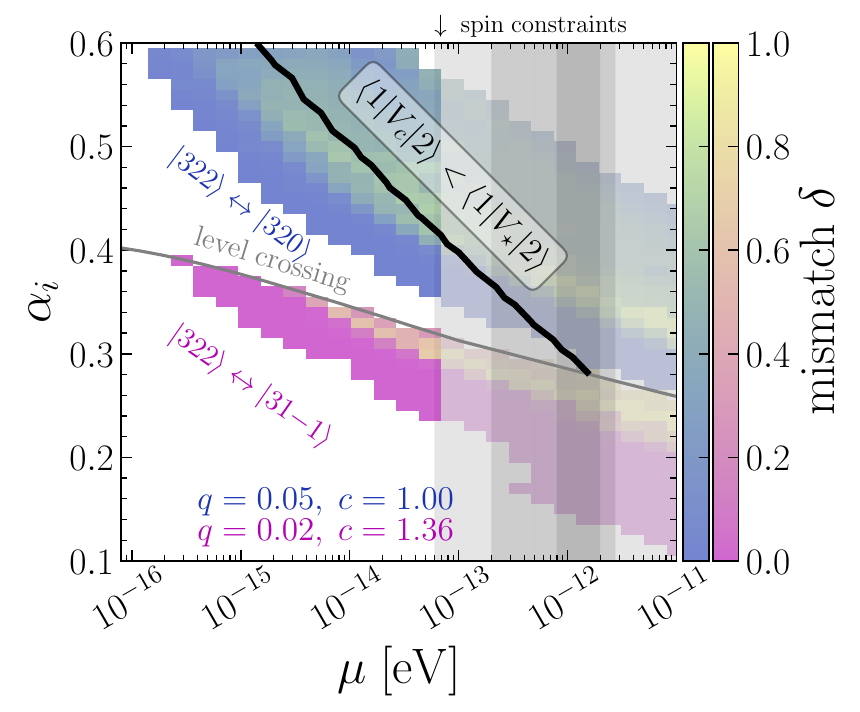}
\caption{Same as Figure~\ref{fig:322_conditions}, but we highlight the region where the off-diagonal self-gravity matrix element is larger than the off-diagonal matrix element of the perturbation due to the secondary. Above the black line, we find $|\langle 1 | V_c | 2\rangle| < |\langle 1 | V_\star |2 \rangle|$. We see that, for the region where the mismatch is large, our treatment of ignoring off-diagonal self-gravity can be justified. Note that $|\langle 1 | V_c | 2 \rangle| < |\langle 1 | V_\star | 2 \rangle|$ during the entire period of the fine resonance. The gray line shows the values of $(\mu,\alpha_i)$ where level crossing between $|322\rangle$ and $|320\rangle$ happens.}
    \label{fig:perturbativity}
\end{figure}

In Figure~\ref{fig:perturbativity}, we numerically check the relative size of $\langle 1| V_\star | 2\rangle$ and $\langle 1 | V_c | 2\rangle$. We obtain the orbital evolution by solving the system  numerically without $\langle 1 | V_c | 2 \rangle$ as before and use the numerical results to compute the relative size of $\langle 1 | V_\star | 2\rangle$ and $\langle 1 | V_c | 2 \rangle$. For the hyperfine $|322\rangle \leftrightarrow |320\rangle$ transition, the off-diagonal element of $V_c$ could be greater than that of $V_\star$  in the region below the thick black line. This region does not overlap much with the parameter space where the mismatch is relevant for detectability $\delta = 1-F > 0.16$. For the fine transition, the off-diagonal self-gravity is always smaller than that of secondary object for the entire evolution. This results suggests that the off-diagonal self-gravity does not play a significant role, especially for the parameter space where the mismatch is non-negligible. In Figure \ref{fig:summary} we show the small region of parameter space where
$|\langle 1 | V_\star | 2\rangle| < |\langle 1 | V_c | 2\rangle|$ with a lighter blue shading.

\subsection{Other effects}
In this work, we neglect environmental effects such as the presence of an accretion disk. An accretion disk can significantly alter the formation history of the superradiance cloud, as the central BH changes its mass at a non-negligible rate \cite{Brito:2014wla}. The main focus of our work is on BHs with masses in the range $10^3 - 10^5\ M_\odot$, as evident from Figure~\ref{fig:summary}. These objects, known as intermediate-mass black holes, are particularly elusive. So far, there have been hints of their role as ultra-luminous X-ray (ULX) sources \cite{Caballero-Garcia:2018mwd}. However, no consensus on the accretion properties of these sources has been reached due to the lack of direct imaging of the disks \cite{Caballero-Garcia:2018mwd}. Moreover, the measurement of the BH mass in ULXs depends on assumptions about the structure of the accretion disk itself, further complicating the search. Therefore, we leave the investigation of possible accretion disk effects on our system to future work.

We neglect phenomena such as boson accretion by the secondary and the dynamical friction experienced by the secondary as it moves through the superradiance cloud. The former effect causes small time-dependent variations in the masses of both the secondary and the cloud, and is subdominant compared to the latter \cite{Baumann:2021fkf}. The latter effect, dynamical friction, is related to the ionization of the boson cloud and has been studied in the literature \cite{Baumann:2021fkf,Tomaselli:2023ysb}. Ionization is a dissipative effect, in the sense that it contributes to the depletion of the cloud once the motion of the secondary object excites bound-free transitions in the cloud. However, ionization has no impact on the orbital dynamics for the binary separations relevant to our work. For the values of $\alpha$ and the resonances we consider, the relevant orbital separations are always $r_\star/r_B > 100$, i.e. much larger than the separations at which ionization is a sizable effect $r_\star/r_B<20$ \cite{Baumann:2021fkf,Tomaselli:2023ysb}.

\section{Conclusion}\label{sec:conclude}
We investigated the implications of the self-gravity of a superradiance cloud for the orbital evolution of the binary system. We showed that self-gravity changes the energy spectrum of the cloud in a density-dependent way and that it introduces a crossing of the $|322\rangle$ and $|320\rangle$ states. We studied the implications of these findings for resonant transitions of the cloud, concentrating on a quasi-circular, prograde and equatorial orbit. Without level crossing, in most cases the cloud is depleted entirely when the orbit enters the resonance $|322\rangle \leftrightarrow |320\rangle$. In contrast, with level crossing, this hyperfine resonance cannot be activated by a prograde orbit in a significant region of parameter space. As a result, the orbit can explore the inner part of the system, potentially triggering the fine resonance $|322\rangle \leftrightarrow |31-1\rangle$. 

We also investigated the observational implications in the context of the future space-borne interferometer LISA. We identified two disjoint regions in the parameter space where LISA can directly probe a GW signal from a binary in a cloud of ultralight bosons. These two disjoint regions are due to the level crossing behavior induced by self-gravity. We found that LISA could probe ultralight bosons in the unconstrained mass range $10^{-15}\,{\rm eV} - 10^{-13}\eV$. Combined with other proposals to probe dense wave dark matter environment around black holes, e.g.~\cite{Kim:2022mdj, Kim:2024rgf}, LISA is expected to probe a wide mass range of ultralight new physics. This is also complementary to other proposed/existing searches, e.g. using black hole spin measurements \cite{Khalaf:2024nwc}, continuous gravitational wave searches \cite{KAGRA:2022osp}, and the motion of S2 stars and their spectroscopy around Sgr A$^*$~\cite{GRAVITY:2023cjt, Bai:2025yxm}.

While promising, our results are limited in several ways. Our analysis remains at the non-relativistic Newtonian level. Although we have shown in a restricted fashion that the relativistic corrections would not qualitatively change the conclusion drawn in the work, a more careful numerical simulation is required to fully determine the detectability of the cloud in LISA for a relatively large $\alpha$. In fact, a recent work by May et al~\cite{May:2024npn} investigated the effects of self-gravity on the continuous gravitational wave emission from boson clouds using numerical simulations. It would be interesting to perform a numerical simulation of a binary system with a cloud, and check if the  numerical results would match with the non-relativistic predictions presented in this work.

In addition, the fitting factor computation involves the simplest vacuum waveform characterized by the chirp mass and the reference frequency. We do not include any other environmental effects, e.g. accretion disk, and post-Newtonian corrections for $h(\boldsymbol \theta_v)$, which will enhance the expressibility of the vacuum waveform, and hence decrease the mismatch of the two waveforms. It would be interesting to include the environmental effects such as those discussed in Ref.~\cite{Cole:2022yzw, Spieksma:2024voy} and at the same time post-Newtonian corrections to repeat the fitting factor computation. We leave this for future study. 

Finally, we considered only a quasi-circular, prograde, equatorial orbit for simplicity. Recent works by Bo{\v s}kovi{\'c} et al~\cite{Boskovic:2024fga} and Tomaselli et al~\cite{Tomaselli:2024bdd, Tomaselli:2024dbw} investigated the impact of orbits with nonvanishing eccentricity and inclination. In particular, the authors of Refs.~\cite{Boskovic:2024fga, Tomaselli:2024dbw} found that the existence of the cloud can leave an interesting signature in the eccentricity distribution of black hole binaries. Extending the present analysis to include eccentricity and inclination as well as possibilities of retrograde orbits could be another interesting future direction. 

\acknowledgements
We thank Aleksandr Chatrchyan, Majed Khalaf, Rotem Ovadia and Ofri Telem for useful discussions. We especially thank Mateja Bo{\v s}kovi{\' c}, Matthias Koschnitzke, and Rafael Porto for discussions and critical comments on the earlier version of this manuscript. We also thank Mattia Arundine for the collaboration and useful discussions in a related project, and Giovanni Maria Tomaselli for useful comments and suggestions. The work of HK is supported by the Deutsche Forschungsgemeinschaft under Germany’s Excellence Strategy - EXC 2121 Quantum Universe - 390833306. The work of AL is supported by an ERC STG grant (``Light-Dark'', grant No. 101040019) and partially by the BSF grant No. 2020220. 
This project has received funding from the European Research Council (ERC) under the European Union’s Horizon Europe research and innovation programme (grant agreement No. 101040019). Views and opinions expressed are however those of the author(s) only and do not necessarily reflect those of the European Union. The European Union cannot be held responsible for them. AL is grateful to the Azrieli Foundation for the award of an Azrieli Fellowship.

\appendix
\section{Equations}\label{app:comoving}
We justify the set of equations we use to investigate the binary system with a boson cloud. We first specify the metric of a rotating  black hole deformed by external matter distribution (Appendix~\ref{app:tidal_pert}), derive the equations governing the internal dynamics of the cloud (Appendix~\ref{app:cloud}), and obtain the angular momentum balance equation and the evolution equations for the black hole and cloud mass (Appendix~\ref{app:balance}).

\subsection{Deformed Metric}\label{app:tidal_pert}
We begin with the metric of an isolated rotating black hole. The Kerr metric is given by
\begin{align}
\!\!\! ds^2 
= 
- \Big( 1 - \frac{ r_s \bar r}{\bar \rho^2}\Big) d\bar t^2
-  \frac{2 a r_s \bar r \sin^2 \bar \theta}{\bar \rho^2} d\bar t d\bar\phi 
+ \frac{\bar \rho^2}{\bar \Delta} d\bar r^2
\nonumber\\
+ \bar \rho^2 d\bar \theta^2
+
\frac{ (\bar r^2 + a^2)^2 - a^2 \Delta \sin^2\bar\theta }{\bar \rho^2} \sin^2\bar \theta d\bar \phi^2
\end{align}
in the Boyer-Lindquist coordinates $\bar x^\mu= (\bar t, \bar r, \bar \theta, \bar \phi)$.
Here $r_s = 2 GM$, $\bar \Delta = \bar r^2 - r_s \bar r + a^2$, $\bar \rho^2 = \bar r^2 + a^2 \cos^2 \bar \theta$, $a = J/M$ and $M$ are the black hole spin and mass, respectively. Since we are interested in the dynamics after the cloud is saturated, the dimensionless spin parameter is $a_* = a / G M = {\cal O}(\alpha)$ and therefore the metric might be expanded to the linear order in the spin parameter assuming $\alpha <1$. We find
\begin{align}
ds^2 \approx
- f d\bar t^2 
+ f^{-1} d\bar r^2 
+ \bar r^2 d\bar \Omega^2
- \frac{2 a r_s \sin^2\theta}{\bar r}  d\bar t d \bar \phi 
\label{isolated}
\end{align}
where $\bar \rho = \bar r$, $f = 1 - r_s /\bar r$, and $d\bar \Omega^2 = d\bar \theta^2 + \sin^2\bar \theta d\bar \phi^2$. 

The above metric is incomplete. It is valid for an isolated black hole, while we consider a system where the black hole is surrounded by another compact object and the boson cloud. The external matter distribution deforms the metric around the rotating black hole, and the resulting metric deviates from the isolated one \eqref{isolated}. 

\begin{figure}
\centering
\includegraphics[width=0.4\textwidth]{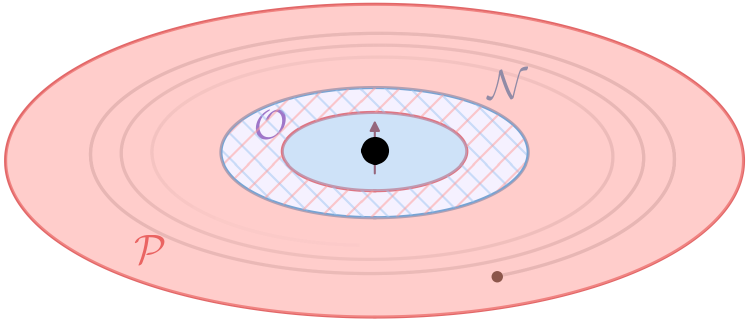}
\caption{A schematic figure describing the matching process to find a tidally deformed metric around the black hole. The region ${\cal N}$ surrounds a rotating black hole, and it is assumed to be vacuum. The region ${\cal P}$ denotes a region where the gravity is weak such that one can perform a post-Newtonian expansion. The boundaries of the region ${\cal N}$ and ${\cal P}$ are represented by solid blue and red lines, respectively. We assume that there exists an overlap region ${\cal O}$ where the metric expanded in ${\cal N}$ and in ${\cal P}$ are both valid. By matching these two metrics, the tidal tensors are determined as a function of external matters.}
\label{fig:schematic}
\end{figure}

Such a tidally deformed metric can be found by following a few steps. First, we consider a deformed metric around a fiducial worldline $\gamma$, along which a rotating black hole is located. Its neighborhood is chosen such that it includes no external matter\footnote{Note that for the largest values of $\alpha$ considered in this work, part of the cloud could enter the ${\cal N}$ region. In this case a metric
reconstruction would be required, including also the cloud on top of the perturbation given by the secondary.}. This region is denoted as ${\cal N}$ in Figure~\ref{fig:schematic}. The metric in this neighborhood is then described by the symmetric and trace-free  electric ${\cal E}_{ij} = R_{0i0j}$ and magnetic tidal tensors ${\cal B}_{ij} = \frac{1}{2} \eps_{i}^{\;kl} R_{0jkl}$ and their derivatives evaluated on $\gamma$~\cite{Thorne:1984mz, Zhang:1986cpa}. These tidal tensors are unspecified at this level. 

The tidal tensors are determined by a matching procedure. We perform a post-Newtonian expansion of the metric in the region ${\cal P}$, where gravity is sufficiently weak. This region includes external matter. We assume that there exists an overlap region ${\cal O}$ where the two metrics --- the one expanded in ${\cal N}$ and the other in ${\cal P}$ --- are both valid. By matching these two metric in ${\cal O}$, one can determine the tidal tensors and its derivative as a function of the external matter distribution. In what follows, we sketch the matching procedure at the Newtonian level. The matching at the post-Newtonian level for non-rotating and slowly-rotating black hole is carried out in detail in Refs.~\cite{Taylor:2008xy, Poisson:2014gka}. See also Ref.~\cite{2014grav.book.....P} for a pedagogical discussion. 

The metric in the neighborhood ${\cal N}$ may be given by~\cite{Zhang:1986cpa, Poisson:2005pi}
\begin{align}
g_{\bar 0 \bar 0}
& \approx 
- 1 + \frac{r_s}{\bar r_h} 
- \sum_{\ell=2}^\infty \frac{2}{\ell(\ell-1)} e_\ell(r) {\cal E}_L \bar x^L 
+ \cdots
\label{00_nearBH}
\\
g_{\bar i \bar j}
&\approx
\delta_{\bar i \bar j }
\label{ij_nearBH}
\end{align}
where $\bar x^L = \bar x^{i_1} \bar x^{i_2} \cdots \bar x^{i_\ell}$, ${\cal E}_L = {\cal E}_{i_1i_2 \cdots i_\ell}$, $\Omega^i = (\sin\theta \cos\phi, \sin\theta \sin\phi, \cos\theta)^i$, and $\lim_{r\to\infty}e_\ell(r) =1$. The metric is expanded at the Newtonian level. Only linear terms of the tidal tensor are kept. In addition, the metric is expressed in Cartesian harmonic coordinates,
$$
\bar x^i 
= (\bar r - G M) \Omega^i 
\equiv \bar r_h \Omega^i. 
$$
The above metric satisfies the harmonic gauge condition, $\partial_\mu (\sqrt{-g} g^{\mu\nu} ) = 0$. The metric in the harmonic gauge is identical to \eqref{isolated} at the Newtonian level. The difference occurs at the post-Newtonian level.

This form of the metric is an interpolation between the metric of an isolated black hole \eqref{isolated} and a tidally deformed metric around a fiducial worldline $\gamma$ without a black hole, where the latter is obtained by Zhang~\cite{Zhang:1986cpa}. With the boundary condition $\lim_{r\to\infty}e_\ell(r) = 1$, one recovers Zhang's metric in the asymptotic region. The detailed form of $e_\ell(r)$ is obtained in Refs.~\cite{Poisson:2005pi, Poisson:2014gka} for non-rotating and rotating black holes by solving the vacuum field equation. For our purpose, the detailed form of $e_\ell(r)$ is unimportant. 

The tidal tensors can be found by matching \eqref{00_nearBH}--\eqref{ij_nearBH} with the post-Newtonian expansion in the overlap region. At the Newtonian level, one finds the metric  in the region ${\cal P}$ as
\begin{align}
g_{00} 
&= - 1 + 2 U(t, \boldsymbol x), 
\\
g_{ij}
&= \delta_{ij}. 
\end{align}
The potentials are given by
\begin{align}
U(t, x)
&=
\frac{G M}{|\boldsymbol x - \boldsymbol z(t)|} 
+ U^{\rm ext}(t, x) , 
\end{align}
where $z^i(t)$ is the position of the black hole in this coordinate system and $U^{\rm ext}(t,x)$ is the potential due to the external matter. We model the rotating black hole as an object with a monopole mass distribution, which is justified by the end result of the matching procedure~\cite{Taylor:2008xy, Poisson:2014gka}. The coordinate system $(t, x^i)$ in which this metric is obtained is the barycenter coordinate system, which differs from the BH comoving coordinate $(\bar t, \bar x^i)$. For the harmonic gauge condition, one also requires $\partial_t U + \partial_j U^j = 0$. 

For the matching, we perform a coordinate transformation to convert the barycenter coordinate to the comoving coordinate, while maintaining harmonic gauge condition and post-Newtonian ordering. Such coordinate transformation is known~\cite{Racine:2004hs, 2014grav.book.....P}: 
\begin{align}
t &= \bar t + \alpha(\bar t, \bar x^i) + \cdots
\label{transformation_time}
\\
x^i &= \bar x^i + z^i(\bar t) + \cdots
\end{align}
where ellipsis denote post-Newtonian corrections. Here~\cite{2014grav.book.....P}\begin{align}
\alpha(\bar t, \bar x^i) = A(\bar t) + v_i(\bar t) \bar x^i,
\end{align}
with $v^i = \dot r^i = d z^i/d\bar t$. The function $A(\bar t)$ is arbitrary at this point, and will be determined by the matching process. Under this transformation, the metric is written as
\begin{align}
g_{\bar 0 \bar 0} 
&= - 1 + 2 \bar U( \bar t, \bar{\boldsymbol x}) , 
\\
g_{\bar i \bar j}
&= \delta_{i j}. 
\end{align}
Each potential is given by 
\begin{align}
\bar U ( \bar t , \bar x)
=& 
\hat U(\bar t , \bar x)
- \dot{A} + \frac{1}{2} v^2 - a_i \bar x^i
\nonumber\\
=&
\frac{G M}{\bar r_h} 
+ \big( \hat U^{\rm ext} - \dot A + v^2/2 \big)
\nonumber\\
&
+ \big( \partial_i \hat U^{\rm ext} - a_i \big) \cdot \bar x^i
+ \sum_{\ell=2}^\infty \frac{\bar x^L}{\ell!}  \partial_L \hat U^{\rm ext}
\end{align}
where $\hat U(\hat t,\bar x) = U(\bar t, \bar x + z)$ and $a_i (\bar t) = \dot v_i (\bar t) = d v_i (\bar t) / d \bar t$. The external potential is expanded around the worldline of the rotating black hole $\bar x = 0$. 

Since the two metrics are given in the same coordinate system in the same gauge, we can finally match them in the overlap region. The matching of the $1/\bar r_h$-term in $g_{\bar 0\bar0}$ justifies the treatment of the black hole as a monopole in the Newtonian expansion of the metric. Furthermore, by matching terms at each order of $\bar x^L$, we find 
\begin{align}
\dot{A} ( \bar t ) &= \hat U^{\rm ext}(\bar t , 0 ) + v^2/2 , 
\\
a_i(\bar t) &= \partial_{i} \hat U^{\rm ext}(\bar t , 0) , 
\\
{\cal E}_L(\bar t)
&= - \frac{1}{(\ell-2)!}\partial_L \hat U^{\rm ext}(\bar t , 0). 
\end{align}
The second line is nothing but the Newtonian equation of motion for the rotating black hole. This matching determines the function $A(\bar t)$ as well as the tidal tensors ${\cal E}_L$. The tidally deformed black hole metric in the comoving coordinate system can be therefore summarized as
\begin{align}
g_{\bar 0 \bar 0}
&\approx - 1 + \frac{r_s}{\bar r_h} 
+ \sum_{\ell=2}^\infty \frac{2}{\ell!} \bar x^L \partial_L \hat  U^{\rm ext} , 
\label{final_0PN_0}
\\
g_{\bar i \bar j}
& \approx \delta_{\bar i \bar j}. 
\label{final_0PN_1}
\end{align}
In the time-time component of the metric, the term $-1 + r_s/\bar r_h$ will provide a $1/r$-potential for the gravitational atom, while the rest will provide the Hamiltonian perturbations $V_\star$ and $V_c$ discussed in the main text. At the post-Newtonian level, the tidal perturbation in the time-space component $g_{\bar t \bar i}$ provides a correction to the energy spectrum from the angular momentum of the cloud. From the power counting, we can already expect that such corrections will be $(v/c)^2 \sim \alpha^2$ suppressed compared to the self-gravity correction arising from $g_{\bar 0 \bar 0}$.

\subsection{Cloud Equation}\label{app:cloud}
We first derive the Schr{\"o}dinger equation in the comoving coordinates discussed in the previous section. We sketch the detailed computations of matrix elements. We then introduce a Bloch equation to solve the three-level system. The resulting set of equations is used in the main text to study the behavior of the system around resonances. 

\subsubsection{Schr{\"o}dinger Equation}\label{app:sch-poi}
For the Schr{\"o}dinger equation, we begin with the action of a minimally coupled scalar \eqref{action},
$$
S = \int d^4  x \sqrt{-g}
\left[ 
- \frac{1}{2} g^{ \mu \nu} \partial_{\mu} \phi \partial_{ \nu} \phi 
- \frac{1}{2} \mu^2\phi^2 
\right] .
$$
Expanding the scalar field in the non-relativistic limit $\phi( \bar t , \bar x) = [ e^{-i \mu \bar t} \psi( \bar t , \bar x) + {\rm h.c.} ] / \sqrt{2\mu}$ and using the tidally deformed metric in the harmonic coordinate \eqref{final_0PN_0} -- \eqref{final_0PN_1}, we find the non-relativistic action at the Newtonian level as
\begin{align}
S \approx \int d^4 \bar x \, 
\psi^* \left[
i \partial_{ \bar t }
+ \frac{\nabla^2}{2\mu}
- V
\right] \psi,
\end{align}
where $V = \mu (1 + g^{00}) / 2$. We have assumed that $| \dot{V} / V | \ll \mu$. Additionally, we have ignored terms like $ (V/\mu) (i \psi^* \partial_t \psi )$ and $(V/\mu) |\nabla\psi|^2/2\mu$ as we remain at the Newtonian level. The Schr{\"o}dinger equation can be read directly from the quadratic action.

The potential can be decomposed into the one due to the rotating black hole $V_1$ and the ones due to the external matter distribution $V_{\rm ext}$,
$$
V = V_1 + V_{\rm ext}.
$$
where $V_1$ arises from $-1+r_s/\bar r_h$ in \eqref{final_0PN_0}, and $V_{\rm ext}$ arises from the rest. 
Together with the kinetic term, the potential $V_1(r)$ constitutes the unperturbed Hamiltonian of the system,
$$H_0 = -\frac{\nabla^2}{2\mu} - \frac{\alpha}{\bar r},$$ 
where $\alpha = GM_1 \mu$ is the gravitational fine structure constant. The system is practically identical to that of the hydrogen atom. 

The external potential consists of two terms: one from the secondary object in the binary and the other from the cloud of ultralight particles. To find the Hamiltonian perturbation, we first note that $\hat U_{\rm ext}(t,\bar{\boldsymbol r})$ of each external matter distribution can be written as
\begin{align}
\hat U_{\rm ext}(t, \bar{\boldsymbol r})
= GM_{\rm ext} \int d^3 \bar r' \,
\frac{\rho_{\rm ext}(\bar t, \bar{\boldsymbol r}')}{| \bar{\boldsymbol r} - \bar{\boldsymbol r}'|},
\end{align}
where $\rho_{\rm ext}(\bar{\boldsymbol r}')$ is the energy density in the comoving coordinate, normalized as $\int d^3 \bar r \, \rho_{\rm ext}(\bar{\boldsymbol r})=1$. For the secondary object, $\rho_{\rm ext}(\bar{\boldsymbol r}) = \delta^{(3)}( \bar{\boldsymbol r} - \bar{\boldsymbol r}_\star )$ with $\bar{\boldsymbol r}_\star = \boldsymbol r_1 - \boldsymbol r_2$ being the separation between the rotating black hole and the secondary object. For the cloud, $\rho(\bar{\boldsymbol r}) = |\psi (\bar{\boldsymbol r})|^2$. From this, the external potential can be found as
\begin{align}
V_{\rm ext} (\bar{\boldsymbol r}) &=
- \mu \sum_{\ell=2}^\infty \frac{\bar x^L}{\ell!} \partial_L \hat U^{\rm ext}_c
\\
&= - GM_{\rm ext}\mu \int d^3 \bar r' \,
\rho_{\rm ext}(\bar{\boldsymbol r}')
\bigg[
\frac{1 }{|\bar{\boldsymbol r} - \bar{\boldsymbol r}'|}
-\frac{1 }{ \bar r'}
- \frac{\bar {\boldsymbol r} \cdot \bar{\boldsymbol r}' }{\bar r'^3}
\bigg] .
\nonumber
\end{align}
The last two terms in the parentheses cancel the monopole and dipole terms in the multipole expansion of $1/|\bar{\boldsymbol r} - \bar{\boldsymbol r}_\star|$ in the limit $\bar r =0$. This justifies the equations we use in the main text for the investigation of the internal dynamics of the cloud. 

\subsubsection{Matrix Element}\label{app:matrix_element}
We use time-dependent perturbation theory to solve the system. In doing so, we compute the matrix element of the perturbation. Note first that the spatial part of the bound-state wave function can be written as a product of the radial wave function and the spherical harmonics
$$
\psi_{n\ell m}(\bar{\boldsymbol r})
= R_{n\ell}(\bar r) Y_{\ell m}(\hat{\boldsymbol r}) ,
$$
where the radial wave function $R_{n\ell}(\bar r)$ is given by
\begin{equation}
\begin{aligned}
R_{n\ell}(\bar r) = 
\sqrt{\Big( \frac{2}{nr_B} \Big)^3 \frac{(n-\ell-1)!}{2n(n+\ell)!}} 
e^{- \bar r / n r_B}
\\
\times
\Big(
\frac{2\bar r}{n r_B}
\Big)^{\ell}
L^{2\ell+1}_{n-\ell-1}
\Big( \frac{2\bar r}{n r_B} \Big) .
\end{aligned}
\end{equation}
Here $r_B =  1 / \mu \alpha$ is the gravitational Bohr radius. Denoting the eigenstate of the unperturbed Hamiltonian as $|i \rangle = | n_i \ell_i m_i \rangle$, the matrix element can be found as
\begin{align}
\langle i | V_{\rm ext}(\bar{\boldsymbol r} ) | j \rangle
&=
- G M_{\rm ext} \mu
\int d^3 \bar r \,
\psi^*_i (\bar{\boldsymbol r})
\psi_j (\bar{\boldsymbol r})
\nonumber\\
&\times
\int d^3 \bar r' 
\rho_{\rm ext} (\bar{\boldsymbol r}')
\left[
\frac{1 }{|\bar{\boldsymbol r} - \bar{\boldsymbol r}'|}
-\frac{1 }{ \bar r'}
- \frac{\bar {\boldsymbol r} \cdot \bar{\boldsymbol r}' }{\bar r'^3}
\right] .
\end{align}

The integral can be decomposed into a radial and angular integral. The quantity in the squared parentheses can be expanded in the spherical harmonics basis,
\begin{align}
&\frac{1 }{|\bar{\boldsymbol r} - \bar{\boldsymbol r}'|}
-\frac{1 }{ \bar r'}
- \frac{\bar {\boldsymbol r} \cdot \bar{\boldsymbol r}' }{\bar r'^3}
\nonumber\\
=& 
\sum_{\ell m}
\frac{4\pi}{2\ell+1}
\bigg(
\frac{r_<^{\ell}}{r_>^{\ell+1}} 
- \frac{\bar r^\ell}{\bar r'^{\ell+1}}\delta_{\ell \leq 1}
\bigg)
Y_{\ell m}^*(\hat{\boldsymbol r}')
Y_{\ell m}(\hat{\boldsymbol r}) , 
\nonumber\\
\equiv &
\sum_{\ell m}
\frac{4\pi}{2\ell+1}
F_{\ell}(\bar r, \bar r')
Y_{\ell m}^*(\hat{\boldsymbol r}')
Y_{\ell m}(\hat{\boldsymbol r}) , 
\end{align}
where $r_> = \max(\bar r, \bar r')$ and $r_< = \min(\bar r,\bar r')$. It is clear that the monopole and dipole vanish in the expansion around $\bar r = 0$. The matrix element can be written as
\begin{align}
\langle i | V_{\rm ext}(\bar{\boldsymbol r} ) | j \rangle
= 
- \frac{ G M_{\rm ext} \mu }{ r_B }
\sum_{\ell m}
\frac{4\pi}{2\ell+1} I^r_{\ell m}(ij) I^\Omega_{\ell m}(ij). 
\end{align}
We introduce the dimensionless integrals 
\begin{align}
I^r_{\ell m}(ij)
&= 
r_B 
\int_0^\infty d \bar r \, \bar r^2 R_i R_j
\int_0^\infty d \bar r' \bar r'^2  F_\ell(\bar r, \bar r') \, \rho_{\ell m}(\bar r')  ,
\label{Ir}
\\
I^\Omega_{\ell m} (ij)
&= \int d\Omega \, 
Y_{\ell_i m_i}^*(\hat r) 
Y_{\ell m} (\hat r) 
Y_{\ell_j m_j} (\hat r) ,
\end{align}
where $\rho_{\rm ext}(\bar{\boldsymbol r}) = \sum_{\ell m} \rho_{\ell m}(\bar r) Y_{\ell m}(\theta,\phi)$ and $R_i = R_{n_i \ell_i}(\bar r)$. The angular integral encodes a set of selection rules, e.g. $m= m_i - m_j$, $|\ell_i - \ell_j| \leq \ell \leq \ell_i + \ell_j$ and $\ell_i + \ell + \ell_j = 2p$ with $p\in {\mathbb Z}$~\cite{Baumann:2018vus}. 

We consider a pointlike particle of mass $M_2$. In this case, the normalized density is $\rho_{\rm ext} (\bar{\boldsymbol r})= \delta^{(3)}(\bar{\boldsymbol r} - \bar{\boldsymbol r}_\star)$, and its spherical harmonics coefficient is $\rho_{\ell m}(\bar r) = \bar r^{-2} \delta(\bar r - \bar r_\star) Y^*_{\ell m}(\hat r_\star)$. 
The radial integral becomes
\begin{align}
I^r_{\ell m} = 
Y^*_{\ell m}(\hat r_\star)
r_B \int_0^\infty d\bar r \, 
\bar r^2 R_i(\bar r) R_j (\bar r) F_\ell(\bar r, \bar r_\star) ,
\label{ir_point}
\end{align}
For a quasi-circular equatorial orbit, the time-dependence can be fully factorized as an exponential, $Y^*_{\ell m} (\hat r_\star) = Y_{\ell m}^*(\pi/2,0) e^{-i m \phi_\star(\bar t)}$ with the orbital phase $\phi_\star(\bar t)$. 
In this case, the matrix element can be written as
\begin{align}
\langle i | V_{\rm ext}(\bar r) | j \rangle
= \gamma_{ij} e^{-i \Delta m_{ij} \phi_\star(\bar t)}
\end{align}
where $\Delta m_{ij} = m_i - m_j$ and 
\begin{align}
\gamma_{ij}
= - \frac{GM_2 \mu}{r_B}
\sum_{\ell \geq | m|}
\frac{4\pi}{2\ell+1} Y_{\ell m}^*(\pi/2,0) I_{\ell m}^{\Omega}(ij)
\nonumber\\
\times
\left[ 
r_B \int_0^\infty d \bar r \,
\bar r^2 R_i R_j F_{\ell}(\bar r, \bar r_\star) 
\right]
\end{align}
Note that the angular integral $I^\Omega_{\ell m}(ij)$ selects $m = \Delta m_{ij} = m_i - m_j$ via the selection rule. The remaining part of the radial integral in \eqref{ir_point} is the same as Eqs.~(3.7) -- (3.9) of Ref.~\cite{Baumann:2018vus}.

\subsubsection{Two-Level System}\label{app:2lv_eq}
Let us now consider a two-level system. We begin with the Schr{\"o}dinger equation
$$
i \dot{\psi}
= ( H_0 + V_\star  + V_c ) \psi , 
$$
where $V_\star$ is the perturbation due to the secondary object in the binary, and $V_c$ is due to the cloud itself. We consider two states denoted as $\{|1 \rangle,  \, |2 \rangle\}$, where $|1\rangle$ represents the dominant cloud state and $|2\rangle$ represents the state that can be resonantly excited from $|1\rangle$ via the perturber. A general state will be written as 
$$
| \psi(\bar t) \rangle = c_1 (\bar t) |1 \rangle + c_2(\bar t) |2\rangle. 
$$ 
We are interested in the evolution of the time-dependent coefficients $c_{1,2}(\bar t)$. We assumed a different magnetic quantum number for each state, i.e. $m_1\neq m_2$ and a quasi-circular and equatorial orbit. 

The Schr{\"o}dinger equation can be written in a matrix form
\begin{align}
i  \dot{\boldsymbol c} _i
= 
[H_0 + V_\star + V_c]_{ij} {\boldsymbol c}_j , 
\end{align}
where the matrix elements are given by
\begin{align}
[H_0]_{ij} 
&= (E_i + i \Gamma_i) \delta_{ij}  ,
\label{H0_element}
\\
[V_\star]_{ij}
&= \gamma_{ij} e^{-i \Delta m_{ij} \phi_\star(t)}  . 
\label{V*_element}
\end{align}
We ignore the diagonal term of $[V_\star]_{ij}$ as it only provides time-independent correction to the energy level, which is parametrically smaller than the self-gravity corrections. In addition, we ignore the off-diagonal matrix elements of the self-gravity term (see Sec.~\ref{sec:off_diagonal}). The Schr{\"o}dinger equation can be explicitly written as 
\begin{align}
i \dot{\boldsymbol c}
= 
\left(
\begin{array}{cc}
E_1 + \Delta E_1 + i \Gamma_1
& \gamma_{12} e^{-i \Delta m_{12} \phi_\star(t)}
\\
\gamma_{12} e^{+i \Delta m_{12} \phi_\star(t)}
& E_2 + \Delta E_2 + i \Gamma_2
\end{array}
\right)
{\boldsymbol c}
\label{hamiltonian_re}
\end{align}

\subsubsection{Three-Level System}\label{app:3lv_eq}
The equation of a two-level system can be generalized to a three-level system. Consider now $\{|1 \rangle, \, |2 \rangle, \, |3\rangle\}$ with an additional spectator state $|3\rangle$ with a large decay width. We are primarily interested in $|3\rangle = |300\rangle$. A generic cloud state can then be written as $|\psi \rangle = c_1(\bar t) |1 \rangle + c_2(\bar t) |2 \rangle + c_3(\bar t) |3 \rangle$. The matrix elements of unperturbed Hamiltonian $[H_0]_{ij}$ and the perturbation due to the secondary object $[V_\star]_{ij}$ are given by the same form as already given in \eqref{H0_element} -- \eqref{V*_element}. The resulting Hamiltonian takes the same form as \eqref{hamiltonian_re}.

This form of Hamiltonian is not particularly convenient for numerical purposes. For the numerical computation, we first eliminate the phases of $[V_\star]_{ij}$ by performing a phase rotation, $c_i \to c_i e^{i\theta_i}$, with
\begin{align}
\theta_i = - m_i \phi_\star(\bar t) + g(\bar t) , 
\end{align}
where $g(\bar t)$ is an arbitrary function. By choosing $2 \dot {g} = (m_1+ m_2) \dot{\phi}_* - (E_1 + E_2)$, we find the the Schr{\"o}dinger equation as
\begin{align}
i \dot{ \boldsymbol c}= 
\left( \begin{array}{ccc}
\frac{\Delta_{12}}{2} + i\Gamma_1
&
\gamma_{12}  
&
\gamma_{13}  
\\
\gamma_{12}  
& 
-\frac{\Delta_{12}}{2} + i\Gamma_2  
& 
\gamma_{23} 
\\
\gamma_{13}
& 
\gamma_{23} 
&
-\frac{\Delta_{13}+\Delta_{23}}{2} + i\Gamma_3
\end{array}
    \right){\boldsymbol c}.
\end{align}
All phases are absorbed into the diagonal elements. Here we introduced $\Delta_{ij} = (E_i+ \Delta E_i) - (E_j + \Delta E_j) - \Delta m_{ij} \Omega$ with the orbital frequency $\Omega(\bar t) = \dot{\phi}_\star(\bar t)$. At the resonance of $\{| i \rangle, \, | j \rangle\}$, $\Delta_{ij} \to 0$. 

Instead of solving for the complex vector $\boldsymbol c$, we solve for elements of the density matrix, i.e.
$$
\rho_{ij} = c_i c_j^*.
$$
In particular, we define
\begin{equation}
\begin{aligned}
\boldsymbol{u}
&= 2 
\big(
{\rm Re} \, \rho_{23},  \,
{\rm Re} \, \rho_{31}, \,
{\rm Re} \, \rho_{12} 
\big)^T , 
\\
\boldsymbol{v}
&= 2 
\big( 
{\rm Im}\, \rho_{32},  \,
{\rm Im}\, \rho_{13}, \,
{\rm Im}\, \rho_{21} 
\big)^T , 
\\
\boldsymbol{w}
&= \sqrt{2} (\rho_{11}, \rho_{22}, \rho_{33})^T . 
\end{aligned}
\end{equation}
We find that the Schr{\"o}dinger equation becomes
\begin{align}
   \left( \begin{array}{c}
 \dot{\boldsymbol u} \\
 \dot{\boldsymbol v} \\
 \dot{\boldsymbol w}
\end{array}
    \right)  = B 
    \left( \begin{array}{c}
{\boldsymbol u} \\
{\boldsymbol v} \\
{\boldsymbol w} \\
\end{array}
    \right)  , 
\label{optial_bloch}
\end{align}
where the $9\times 9$ anti-symmetric matrix $B$ is given by
\begin{widetext}
\renewcommand{\arraystretch}{1.3}
\begin{align}
B
=
\left(
\begin{array}{ccc|ccc|ccc}
\Gamma_{23} 
& 0
& 0
& -\Delta_{23}
& \gamma_{12}
& -\gamma_{13}
& 0
& 0
& 0
\\
0
& \Gamma_{13}
& 0
& -\gamma_{12}
& - \Delta_{31}
& \gamma_{23}
& 0
& 0
& 0 
\\
0
& 0
& \Gamma_{12}
& \gamma_{13}
& -\gamma_{23}
& - \Delta_{12}
& 0
& 0
& 0 
\\ \hline
\Delta_{23}
& \gamma_{12}
& - \gamma_{13}
& \Gamma_{23}
& 0
& 0
& 0 
& - \sqrt{2} \gamma_{23}
& \sqrt{2} \gamma_{23}
\\
-\gamma_{12}
& \Delta_{31}
& \gamma_{23}
& 0
& \Gamma_{13}
& 0 
& \sqrt{2}\gamma_{13}
& 0 
& - \sqrt{2}\gamma_{13}
\\
\gamma_{13}
& - \gamma_{23}
& \Delta_{12}
& 0
& 0 
& \Gamma_{12}
& - \sqrt{2} \gamma_{12}
& \sqrt{2} \gamma_{12}
& 0
\\ \hline
0 
& 0
& 0
& 0 
& - \sqrt{2} \gamma_{13}
& \sqrt{2} \gamma_{12}
& 2 \Gamma_1
& 0 
& 0
\\
0
& 0
& 0
& \sqrt{2} \gamma_{23}
& 0 
& - \sqrt{2} \gamma_{12}
& 0
& 2 \Gamma_2
& 0
\\
0
& 0
& 0
& - \sqrt{2} \gamma_{23}
& \sqrt{2} \gamma_{13}
& 0
& 0
& 0
& 2 \Gamma_3
\end{array}
\right)\ ,
\end{align}
\end{widetext}
where $\Gamma_{ij} = \Gamma_i + \Gamma_j$. This resembles an optical Bloch equation. This equation is simultaneously solved with the equations for the spin and mass of the black hole and the orbital frequency, which will be discussed in the next section.

\subsection{Balance Equation}\label{app:balance}
We introduce a set of equations to investigate the interaction between the cloud and the binary system. In Appendix~\ref{app:ang_mom}, we review the angular momentum balance equation, which is used in the main text to study the backreaction of the cloud onto the binary system and the evolution of the gravitational wave frequency. In Appendix~\ref{app:cloud_mass}, we review the evolution equations for the black hole and the cloud mass.

\subsubsection{Angular Momentum}\label{app:ang_mom}
The angular momentum balance equation is given by
\begin{align}
\frac{d {\pmb J}}{dt} = - \pmb{\cal T} , 
\end{align}
where $\pmb J$ is the total angular momentum of the system and 
$\pmb{\cal T}$ is the gravitational torque. We work in the barycenter coordinate system $(t,\boldsymbol x)$ and assume a quasi-circular equatorial orbit. The spin of the black hole is aligned along the $+\hat z$ direction. Furthermore, we assume that the center-of-mass of the cloud coincides with the rotating black hole.\footnote{When the cloud is composed of a coherent superposition of two states with different parity, the center-of-mass of the cloud in the comoving frame might oscillate around the black hole. Although this could happen in the case of a fine transition $\{|322\rangle, |31-1\rangle\}$, we do not expect a large deviation of the center-of-mass of the cloud from the center of the comoving frame, as the relative occupation number of the $|31-1\rangle$ state remains very small due to its decay.}

Along the $\hat z$-direction, the torque is given by~\cite{2014grav.book.....P}
\begin{align}
{\cal T}_z = 
{\rm sign}(L_o) \frac{32}{5} \frac{\mu_r^2}{M} 
\left( \frac{GM}{r_*} \right)^{7/2} ,
\end{align}
where $M = M_1 + M_2 + M_c$ is the total mass and $\mu_r = (M_1 + M_c) M_2 / M$ is the reduced mass between the secondary object and the black hole-cloud system. The total angular momentum is given by
\begin{align}
J_z = 
L_o + J_c + J_{\rm BH} , 
\end{align}
where $J_{\rm BH} = a_* G M_1^2$ is the spin of the black hole, $J_c = (M_c / \mu) \sum_i m_i |c_i|^2$ is the total angular momentum of the cloud, and $L_o = \pm \mu_r \sqrt{G M r_\star}$ is the orbital angular momentum. The sign of the orbital angular momentum is determined by the orientation of the orbit; it takes the $+$ ($-$) sign for a prograde (retrograde) orbit. 

The angular momentum balance equation can be written as an equation for the orbital frequency. As we assume a quasi-circular orbit, the orbital frequency is given by
\begin{align}
\Omega 
= \sqrt{ \frac{G M}{r_\star^3} } . 
\end{align}
We find that the balance equation can be written as
\begin{align}
\dot{\Omega}
= &
\frac{96}{5} \frac{\mu_r}{M} (GM)^{5/3} \Omega^{11/3}
\nonumber\\
&\pm \frac{3 \Omega^{4/3}}{(GM)^{2/3}} \frac{M_c}{\mu_r}
\sum \frac{m_i}{\mu} 
\left( \frac{d |c_i|^2}{dt} - 2 \Gamma_i |c_i|^2 \right). 
\label{freq_evol}
\end{align}
In deriving the above equation, we use the following angular momentum evolution equation~\cite{Brito:2014wla, Hui:2022sri} 
\begin{align}
\dot{J}_{\rm BH}
\approx - 2 M_c \sum_i \frac{m_i}{\mu} \Gamma_i |c_i|^2,
\end{align}
which is obtained by computing the angular momentum flux across the black hole horizon. The above equation is only approximate. For multiple cloud modes, there exist interference terms in the above expression. Such terms oscillate much faster than the typical time scale at which the black hole spin changes, and hence they can be neglected~\cite{Hui:2022sri}. 

We solve \eqref{optial_bloch} and \eqref{freq_evol} altogether. However, they are given in different coordinate systems; the Bloch equation is given in the comoving coordinate $(\bar t, \bar{\boldsymbol x})$, while the frequency evolution equation is given in the barycenter coordinate system $(t, \boldsymbol x)$. The transformation between the barycenter time and the comoving time is given in \eqref{transformation_time}. The difference of these two time coordinates will give rise to additional terms multiplied by $\alpha(\bar t, \bar x^i)/ c^2$ in \eqref{freq_evol}. As we remain at the Newtonian level, we will ignore such terms and simply take $t = \bar t$ for the numerical computation.

\subsubsection{Mass}\label{app:cloud_mass}
For the discussion in the main text, we use the numerically obtained cloud mass fraction $q_c = M_c / M_1$. We detail how this is obtained in a similar way to the discussion in \cite{Khalaf:2024nwc}.

\begin{figure}
\centering
\includegraphics[width=0.48\textwidth]{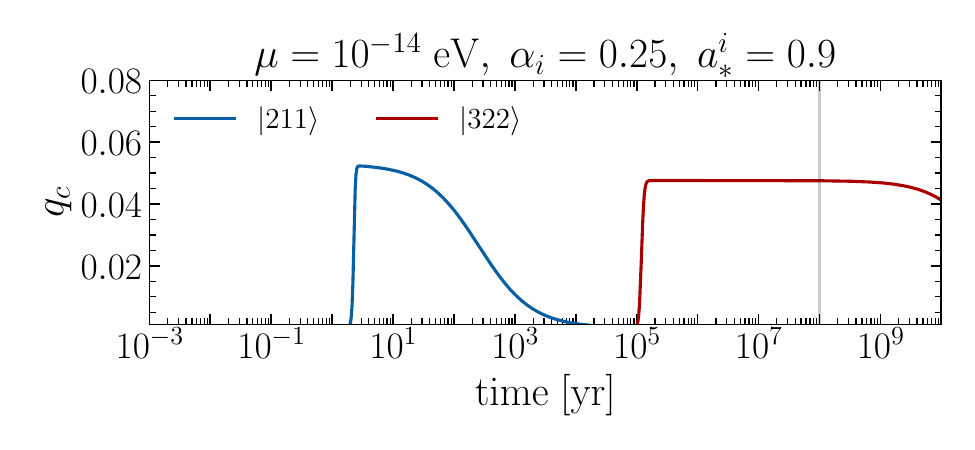}
\caption{The time evolution of the cloud mass for the benchmark point $(\alpha_i, \mu) = (0.25, 10^{-14}\eV)$ shown as a star in Figure~\ref{fig:qc}. The initial spin is set to $a_*^i=0.9$. The blue line shows when the cloud is in the $|211\rangle$ state, while the red line shows when it is in the $|322\rangle$ state. The vertical line corresponds to the reference time 100 Myr at which the cloud is fully saturated. }
\label{fig:qc_max_100Myr_2}
\end{figure}

We solve a set of equations governing the evolution of the black hole mass, the cloud mass, and the spin of the black hole. In particular, we solve
\begin{align}
\frac{d M_1}{ dt } &= - \sum_i 2 \Gamma_i M^c_i ,
\label{dM1dt}
\\
\frac{d M^c_i}{dt} &= 2 \big[ \Gamma_i  - ( \Gamma_{i}^{\rm GW} / \mu ) M^c_i \big] M^c_i,
\\
\frac{da_*}{dt} &= - \sum_i \left( \frac{m_i}{\alpha} - 2 a_* \right) 2\Gamma_i \frac{M^c_i}{M_1} ,
\label{dadt}
\end{align}
where $M_c^i$ is the mass of $i$-th superradiance state, e.g. $i = |211\rangle, \, |322\rangle, \, \cdots$, and $\Gamma_i^{\rm GW}$ is the annihilation rate of the $i$-th cloud state into gravitational waves. The annihilation rates are given as~\cite{Baryakhtar:2020gao}
\begin{align}
\Gamma_{211}^{\rm GW} / \mu
&=  10^{-2} \alpha^{13} (\mu / M_1) ,
\\
\Gamma_{322}^{\rm GW} / \mu
&= (3 \times 10^{-8}) \alpha^{17} (\mu / M_1) .
\end{align}
In Figure~\ref{fig:qc}, we show numerical results on the cloud mass fraction at $t_{\rm sys} = 100\,{\rm Myr}$ in the parameter space $(\alpha, \, \mu)$. For presentation, we choose to show only when $q_c > 10^{-5}$. 

The maximum cloud mass fraction can be analytically estimated. The superradiance instability extracts the angular momentum of the black hole by exponentially producing ultralight particles in a sequential manner. For example, the extraction first takes place through the production of $|211\rangle$ for a relatively small fine structure constant. The $|211\rangle$ cloud saturates when
\begin{align}
\Gamma \propto (m \Omega_+ - \mu )\approx 0 .
\label{sat_condition}
\end{align}
Here $\Omega_+ = a_* / 2r_+$ and $r_\pm = G M_1( 1 \pm \sqrt{1-a_*^2})$. Even after the $|211\rangle$ cloud saturates, further extraction can occur via the production of $|322\rangle$ and states with higher angular momentum. 

Consider now a scenario where the $k$-th dominant superradiant state saturates. The black hole mass at this moment is denoted as $M_1^{(k)}$ and the $k$-th superradiant cloud mass is denoted as $M_c^{(k)}$. Note that $M_1^{(k)} = M_1^{(k-1)} - M_c^{(k)}$. The dimensionless spin parameter at the saturation can be obtained by solving \eqref{sat_condition}. We find
\begin{align}
a_{*}^{(k)} = \frac{4 \alpha^{(k)} / m_k}{1 + 4 (\alpha^{(k)}/ m_k)^2}
\label{sat_a*}
\end{align}
where $\alpha^{(k)}$ is the fine structure constant at the point of the saturation of the $k$-th dominant cloud, and  $m_k$ is the magnetic quantum number of such cloud state. 

Using the definition of the dimensionless spin parameter $a_* = J / GM_1^2$, we can write $a_*^{(k)}$ as
\begin{align}
a_*^{(k)} 
&= \frac{J^{(k)}}{G [M_1^{(k)}]^2}
= \frac{J^{(k-1)} + \Delta J}{G (M_1^{(k-1)} - M_c^{(k)} )^2}
\nonumber\\
&= a_*^{(k-1)} \Big( 1 + \frac{M_c^{(k)}}{M_1^{(k)}} \Big)^2
- \frac{m_k}{\alpha^{(k-1)}} \frac{M_c^{(k)}}{M_1^{(k)}} \Big( 1 + \frac{M_c^{(k)}}{ M_1^{(k)}} \Big)
\end{align}
where we use $\Delta J = - m_k (M_c^{(k)}/\mu)$. Substituting \eqref{sat_a*} into the above equation, using $\alpha^{(k)} = \alpha^{(k-1)} / (1 + M_c^{(k)}/ M_1^{(k)})$, and solving for $q_c^{(k)} = M_c^{(k)}/ M_1^{(k)}$, we find
\begin{align}
q_c^{(k)} 
= - 1 + \frac{1 + \sqrt{1 - 16 (\frac{\alpha^{(k-1)}}{m_k})^2 (1- \frac{a_*^{(k-1)} \alpha^{(k-1)} }{ m_k })^2}}{2 ( 1 - a_*^{(k-1)} \alpha^{(k-1)} / m_k)}
\label{qc_est}
\end{align}
This can be solved iteratively. 
When solving \eqref{qc_est} iteratively, one must carefully choose the dimensionless spin parameter and the fine structure constant at each iteration. When $k$-th superradiant state begins to be produced, it renders the $(k-1)$-th state unstable. The $(k-1)$-th state could subsequently decay back to the black hole, spinning up the black hole and increasing the black hole mass. For the cases considered in this work, the $(k-1)$-th state annihilates into gravitational waves before it can decay to the black hole, and hence, the dimensionless spin parameter and fine structure constant at each iteration are approximately given by the saturated value \eqref{sat_a*}. 
For the maximum mass fraction of the $|211\rangle$ state, the above estimation is identical to the estimation given as (F11) in Ref.~\cite{Baryakhtar:2020gao}.

\section{Mixing-Induced Decay}\label{app:decay}
The mixing between a superradiance state and a rapidly decaying state can significantly change the dynamics of the binary system. To demonstrate this, we consider a two-level system without self-gravity. We consider $\{ |1 \rangle, |3 \rangle\}$ where the spectator state $|3\rangle$ is assumed to have a large decay width, e.g. $|3\rangle = |300\rangle$. 

The Schr{\" o}dinger equation is
$$
i \dot{\boldsymbol c}_i
= 
[H_0 + V_\star]_{ij}
{\boldsymbol c}_j.
$$
For the discussion, we ignore the self-gravity correction to the energy level and also the instability or decay width of the states $|1\rangle$. The unperturbed Hamiltonian is then approximated as $ H_0 = {\rm diag}(E_1 , E_3 + i \Gamma)$. The matrix element for the perturbation is given by $[V_\star]_{ij} = \gamma_{ij} e^{ - i \Delta m_{ij} \phi_\star}$. By performing a diagonal phase rotation, $c_i \to e^{-i E_i t} c_i$, the Schr{\"o}dinger equation can be written as
\begin{align}
i \dot{\boldsymbol c}
&=
\left(
\begin{array}{cc}
0 
& \gamma e^{i \Delta} 
\\
\gamma e^{ - i \Delta} 
& i \Gamma
\end{array}
\right)
{\boldsymbol c} . 
\end{align}
where $\Delta = (E_1 - E_3)t - \Delta m_{13} \phi_\star(t)$ and $\gamma = \gamma_{13}$. The initial condition of the system is $c_1(-\infty) =1$ and $c_3(-\infty) = 0$. 

Solving the Schr{\"o}dinger equation for $c_{3}$ up to ${\cal O}(\gamma)$, we find
\begin{align}
c_3(t)
&\approx - i \int^t dt' \,
\gamma e^{- i \Delta(t')} e^{\Gamma(t - t')} . 
\end{align}
Using the above result, we find the solution for $c_1$ up to ${\cal O}(\gamma^2)$ is 
\begin{align}
\dot{c}_1(t)
&\approx
- \gamma^2 \int^{t} dt' \, e^{ i [ \Delta (t) - \Delta (t')]} e^{\Gamma(t-t')} . 
\end{align}
Most of the contribution arises from $t' \approx t$ as the phase oscillates rapidly. Expanding $\Delta(t') = \Delta(t) + \dot \Delta (t'-t)$ and ignoring the slow time-dependence of $\gamma(t)$, we find that the real part of $\dot{c}_1 / c_1$ is
\begin{align}
{\rm Re} \, \frac{\dot{c}_1}{c_1} 
\approx \frac{\gamma^2 \Gamma}{[(E_1 - E_3) - \Delta m_{13} \Omega]^2 + \Gamma^2}. 
\end{align}
This reproduces \eqref{induced_decay}. While this term arises only in the second order perturbation theory, it can significantly modify the orbital dynamics due to the large decay width $\Gamma$ of the spectator state.

\section{Calculation of the Fitting Factor}\label{app:fitting factor}
We consider a scenario in which the gravitational wave is emitted from a binary system embedded in the cloud. We use the fitting factor to estimate the detectability of the cloud in LISA via gravitational waves observations. The fitting factor is defined as
$$
{\cal F} = {\rm max}_{\boldsymbol \theta_v} 
\frac{
\big( \, 
h(\boldsymbol \theta_{\rm true}) \, \big| \, 
h(\boldsymbol \theta_v) \, 
\big)}{ 
\sqrt{ 
\big( \, 
h(\boldsymbol\theta_{\rm true}) \, \big| \, 
h(\boldsymbol\theta_{\rm true}) \, \big) 
\big( \, 
h(\boldsymbol\theta_v) \, \big| \, 
h(\boldsymbol\theta_v) \, \big) } } \ , 
$$
with the inner product given in Eq.~\eqref{inner_product}. Here $h(\boldsymbol \theta)$ represents the waveform of GWs emitted from the system with the cloud. The inner product is maximized only over a subset of parameters $\boldsymbol \theta_v = \{ f_-, \, {\cal M}_c\}$ as discussed in the main text. The waveform $h(\boldsymbol \theta_v)$ represents the GWs that would have been emitted from the system without the cloud of ultralight particles. The fitting factor hence measures the degree of mismatch between two waveforms that represent GWs emitted from the system with and without the cloud. The extrinsic parameters are already maximized in the above expression. The fitting factor takes a value ${\cal F} \in [0,1]$; ${\cal F}=1$ would indicate that two waveforms are indistinguishable. We detail below the computation of the fitting factor. 

\begin{figure*}[!t]
\centering
\includegraphics[width=0.4\textwidth]{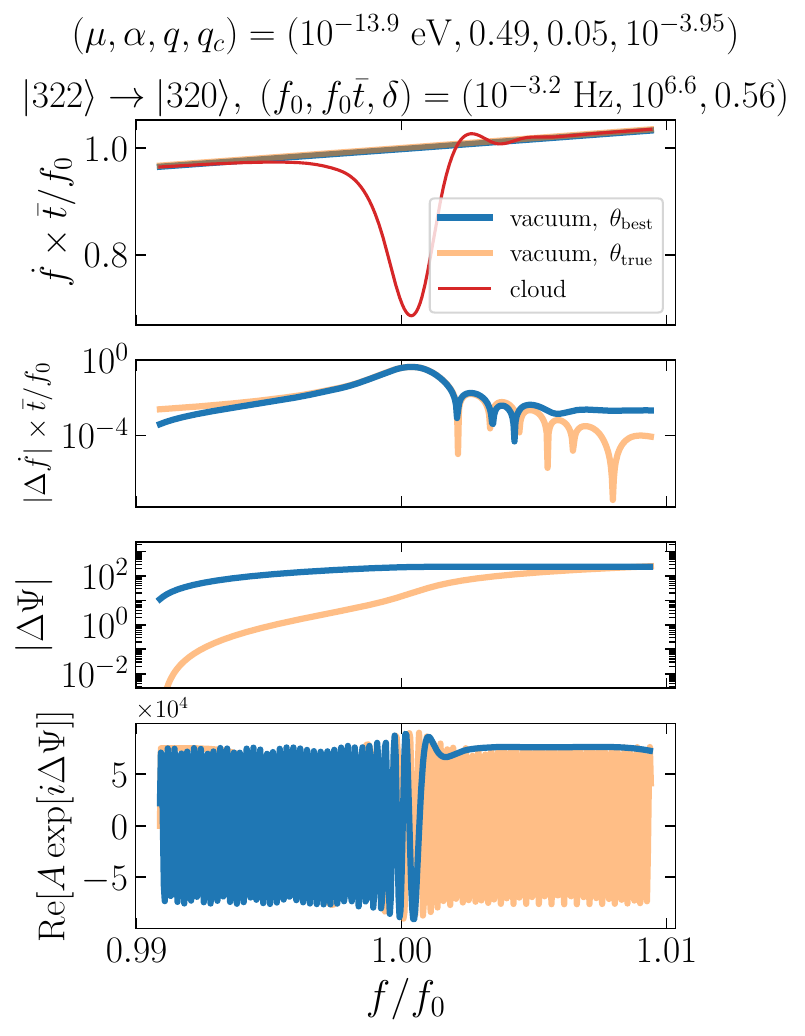}
\includegraphics[width=0.4\textwidth]{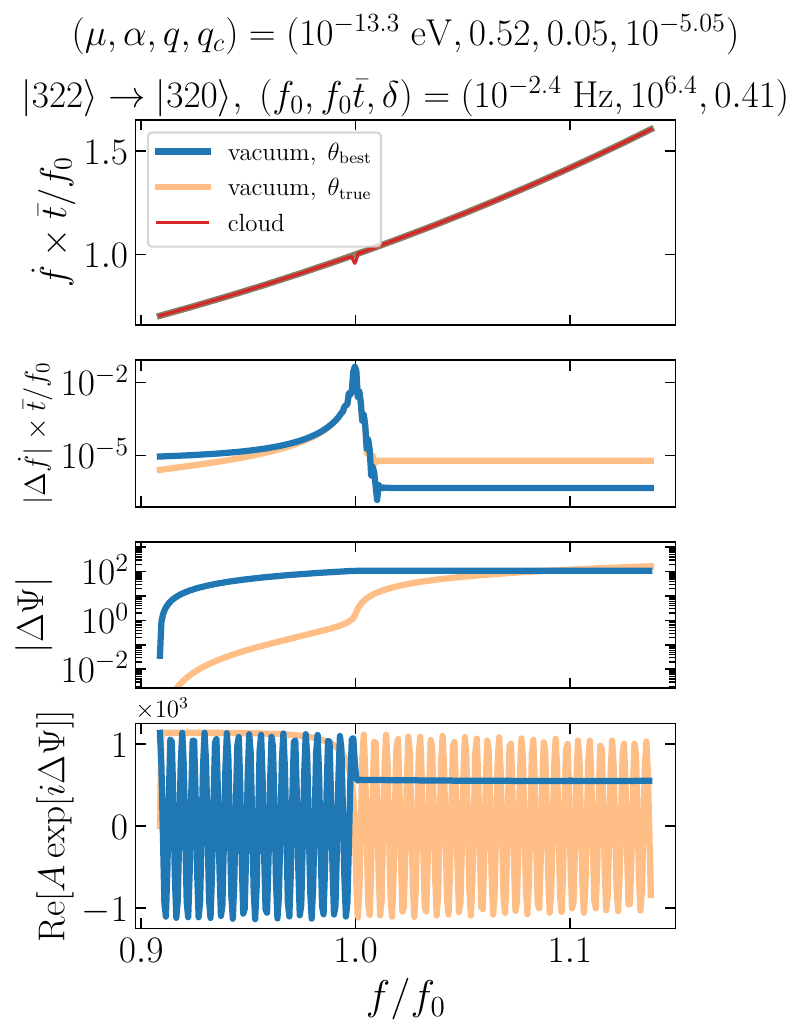}
\includegraphics[width=0.4\textwidth]{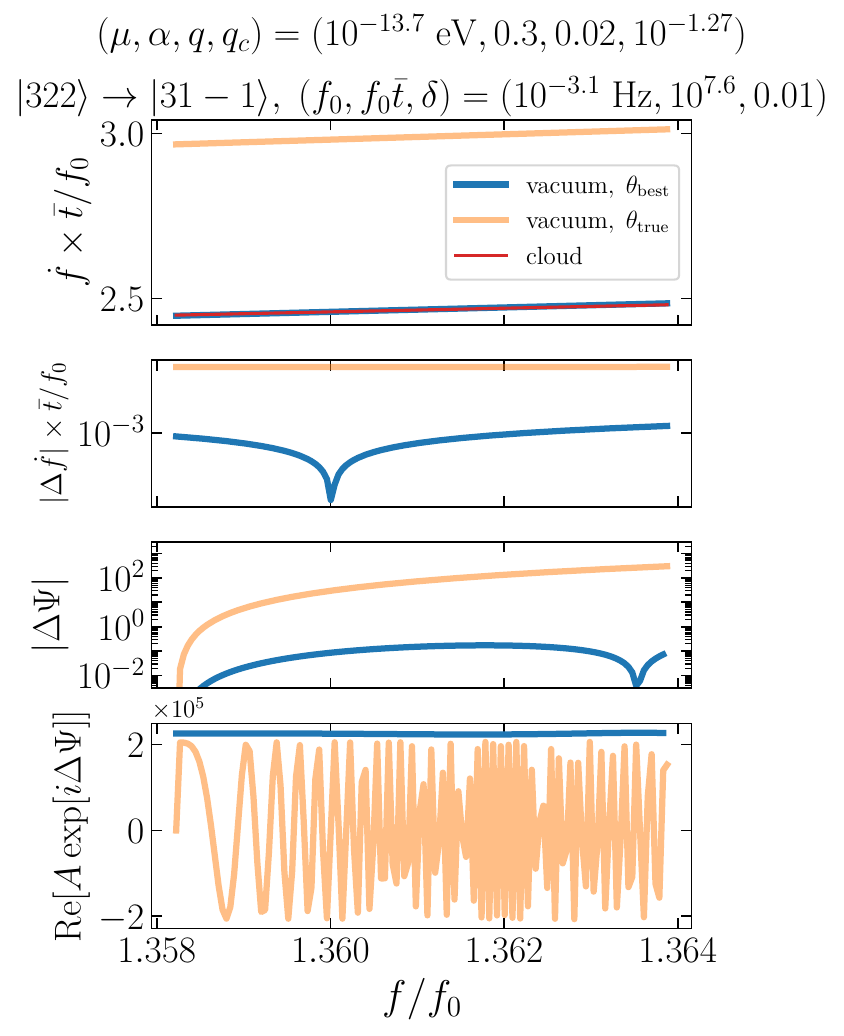}
\includegraphics[width=0.4\textwidth]{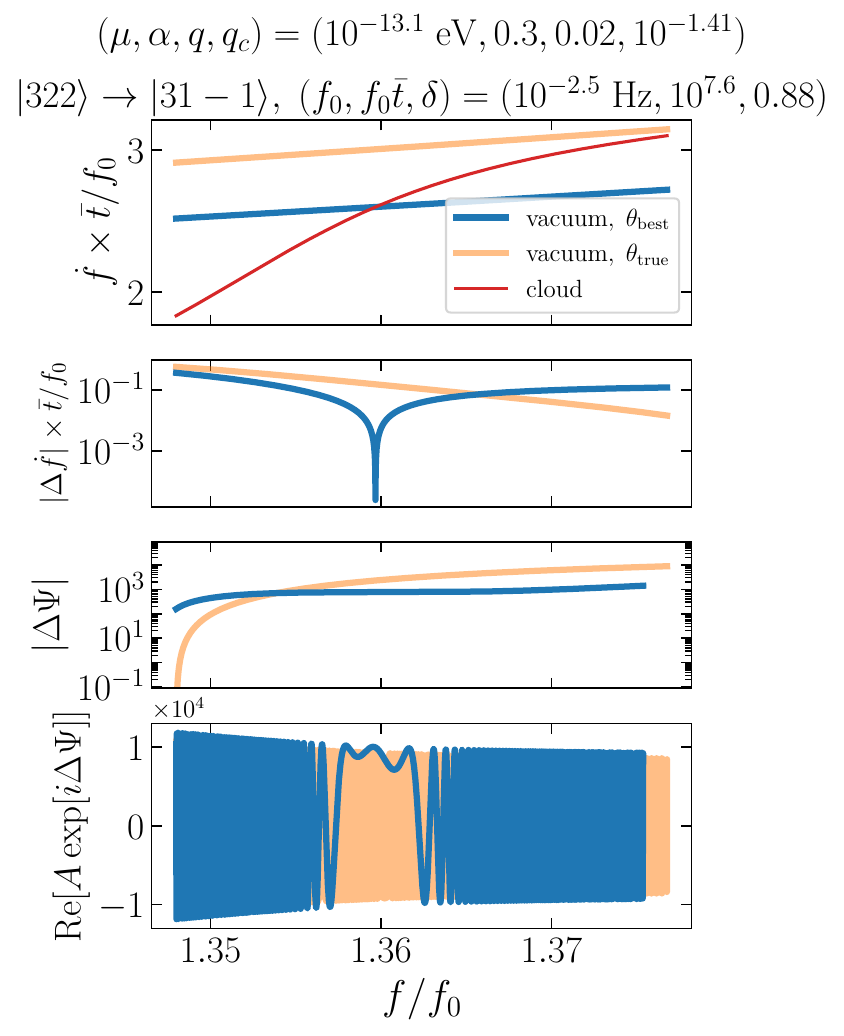}
\caption{   Examples of calculation of the fitting factor. Upper plots show hyperfine transitions aligned with the resonance frequency $c=f(t_0)/(\Omega_0/\pi)=1.0$, while lower plots show fine transitions with $c=1.36$. The other parameters are shown in the plot titles. Note that $f_0 = \Omega_0/\pi$ and $\bar t = [\Omega_0 (d\Omega/dt)^{-1}|_{\Omega=\Omega_0}]$ are computed with true parameters $\boldsymbol \theta_{\rm true}$ at the beginning of numerical evaluation. For each plot, each panel shows, from top to bottom, (i) the frequency evolution $\dot{f} \times \bar t / f(t_0)$, (ii) the difference in the frequency evolution between different hypotheses, (iii) the phase difference $|\Delta \Psi|$, and (iv) the real part of the integrand of the numerator the fitting factor in Eq.~\eqref{eq:F_final}. The blue line presents the result obtained by fitting the vacuum waveform to the true signal, while the orange is the result without any fitting procedure. From the bottom panel of each plot, the mismatch $\delta$ can be estimated by counting the fraction of the frequency range where the integrand is rapidly oscillating. }
\label{fig:F_examples}
\end{figure*}

We assume that the detector operates in the time interval $t\in [t_-,t_+]$ with $t_\pm = t_0\pm T_{\rm obs}/2$, where $t_0$ is an arbitrary reference time, and $T_{\rm obs}$ is the total observational time span. The waveform is given by~\cite{Maggiore:2007ulw}
\begin{align}
h(f) = A \frac{(\pi f)^{2/3}}{  {\dot f}^{1/2} }e^{i\Psi(f)} \Theta(f_+ -f)\Theta(f-f_-) . 
\end{align}
Here $A$ is the frequency independent amplitude of the strain, $\Psi$ is the phase, and $f_\pm$ is the frequency of gravitational wave at the beginning and end of the observational campaign $t_\pm$. The step functions ensure that the strain vanishes when the GW frequency is not in the band $[f_-,f_+]$. The phase can be written as~\cite{Maggiore:2007ulw}
\begin{align}
\Psi(f) 
&= 2 \pi f ( t_* + r ) - \Phi(t_*) -  \vartheta,
\end{align}
where $\vartheta$ is a constant phase factor, $r$ is the distance to the source, and $\Phi(t_*)$ is the phase of gravitational wave at the retarded time $t_*$ defined through $2\pi f = \dot{\Phi}(t_*)$. The time $t_*$ can be interpreted as the time at the source frame when the gravitational wave of frequency $f$ is emitted. For the following discussion, we rewrite the phase as
\begin{align}
\Psi(f)
&= 2\pi f t_{-} + {\cal N}(f) - \tilde\vartheta
\label{Psi_phase}
\end{align}
where $\tilde \vartheta$ is a constant phase, and ${\cal N}$ is defined as
\begin{align}
{\cal N} (f)
&= 2\pi \int^{f}_{f_-} df' \frac{(f - f')}{\dot f'}. 
\label{N_phase}
\end{align}
The main difference between two waveforms arises from the difference in their phase evolution $\dot{f}$ inside the integral in ${\cal N}$ in Eq~\eqref{N_phase}.

When computing the fitting factor, we assume gravitational waves emitted from the cloud-binary system as a true signal. We also assume $f(t_0) \simeq  (\Omega_0/\pi) $; the gravitational waves emitted at the resonance enter the detector at the middle of the observational campaign. The entry and exit frequencies $f_\pm$ can be computed at each point in the parameter space $(\alpha,\mu,q_c)$. We then repeat the computation for different choices of GW frequencies that enter the detector in the middle of the observational campaign $f(t_0) \simeq  c (\Omega_0/f)$ with $ c \neq1$.

The inner product of the waveforms is 
\begin{align}
\big( \, 
h(\boldsymbol \theta_v) \, \big| \, 
h(\boldsymbol \theta_v) \, \big)
&= 4 A_v^2 
\int _{f_-^v}^{f_+^v}df \frac{(\pi f)^{4/3}}{S_n(f)  }
\frac{1}{ \dot f_v } , 
\\
\big( \, 
h(\boldsymbol \theta_{\rm true}) \, \big| \, 
h(\boldsymbol \theta_{\rm true}) \, \big)
&= 
4 A_{\rm true}^2 
\int _{f_-^c}^{f_+^c}df \frac{(\pi f)^{4/3}}{S_n(f) }
\frac{1}{ \dot f_c } , 
\end{align}
where $\dot{f}_c$ and $\dot{f}_v$ are the frequency evolution with and without the cloud, and $S_n(f)$ is the noise power spectral density in the strain unit. The inner product between the two templates is
\begin{align}
\big( \, 
h(\boldsymbol \theta_{\rm true}) \, \big| \,
h(\boldsymbol \theta_v) \, \big) 
= 
4 A_{\rm true} A_v {\rm Re}\int _{f_-}^{f_+}df \frac{(\pi f)^{4/3}}{S_n(f)} \frac{e^{i\Delta \Psi}}{(\dot f_v \dot f_c)^{1/2}}
\end{align}
where $f_- = \max(f_-^v, f_-^c)$, $f_+ = \min(f_+^v, f_+^c)$, and $\Delta \Psi = \Psi(\boldsymbol \theta_v) - \Psi(\boldsymbol \theta_{\rm true})$. The vacuum waveform is governed by two parameters ${\boldsymbol \theta}_v = \{f_-^v, {\cal M}_c \}$, while the other parameters are taken as $q_c \to 0$, $\alpha\to \infty$, and $\mu \to \infty$. The exit frequency $f_+^v$ is determined by the other two parameters, $f_+^v  = f_+^v (f_-^v, {\cal M}_c)$.

Combining these inner products, we find the fitting factor as
\begin{align}\label{eq:F_final}
{\cal F} = \max_{\boldsymbol \theta_v} 
\frac{\bigg|\int_{f_-}^{f_+} df \frac{(\pi f)^{4/3}}{S_n(f)} \frac{e^{ i \Delta {\cal N}}}{(\dot f_v \dot f_c)^{1/2}}\bigg|}{\bigg[\int _{f_-^v}^{f_+^v}df \, \frac{(\pi f)^{4/3}}{S_n(f)  } \frac{1}{\dot f_v } \bigg]^{\frac12}\bigg[\int _{f_-^c}^{f_+^c}df \frac{(\pi f)^{4/3}}{S_n(f)} \frac{1}{\dot f_c } \bigg]^{\frac12}}.
\end{align}
where ${\cal N} = {\cal N}(\boldsymbol \theta_v) - {\cal N}(\boldsymbol \theta_{\rm true})$. The absolute value of the numerator appears after one maximizes the extrinsic parameter $\Delta\tilde\vartheta$.

The integral in the numerator of \eqref{eq:F_final} is computed via a fast Fourier transform. The number of frequency bins $N_f$ is decided based on the minimum value of the phase difference found before a scan of the parameter space. For a meaningful evaluation of the oscillating integrand, we require $N_f > 2|\min[\Delta {\cal N}(f_-^c,{\cal M}_c^{\rm true}),\Delta {\cal N}(f_-^c,{\cal M}_c^{0})]|$, with ${\cal M}_c^{0}$ defined such that $ \dot f_c(f(t_0)) = \dot f_v(f(t_0))|_{ {\cal M}_c = {\cal M}_c^0}$. The maximization over ${\boldsymbol \theta}_v$ is achieved via the Nelder-Mead algorithm in the \texttt{scipy.optimize} library. The algorithm takes an initial guess of $\boldsymbol \theta_v$ as input. To search for the global maximum efficiently, we run the algorithm from $10^3$ different initial points. The initial point for $f_-^v$ is fixed to $f_-^c$, while the initial point for ${\cal M}_c$ is sampled from the normal distribution centered around either ${\cal M}_c^{\rm true}$ or  ${\cal M}_c^{0}$, depending on which one provides the lowest $\Delta {\cal N}$. The standard deviation of the normal distribution is varied a few times to ensure a good coverage of the parameter space of $\boldsymbol \theta_v$.

We present some examples of the fitting factor calculation in Figure~\ref{fig:F_examples}. We show the behavior of frequency evolution, phase difference, and other related quantities for the vacuum waveform with the best-fit parameter $\boldsymbol \theta_{\rm best}$ obtained from the procedure described above and with $\boldsymbol \theta_{\rm true}$ without any fitting procedure. The upper plots show examples for the hyperfine transition with $\alpha=0.5$, while the lower plots show the examples of fine transitions with $\alpha=0.3$. Other parameters chosen for this numerical analysis are shown in the plot title. 

Both examples of the hyperfine transition exhibit a clear sign of resonance behavior. While the frequency evolution before and after the resonance is similar to the vacuum evolution, the resonance introduces a sudden dephasing of gravitational waves before or after the resonance. Even if the parameters $\boldsymbol \theta_v$ are chosen so that it fits the true signal after (before) the resonance, such a vacuum waveform still leads to a large dephasing due to the mismatch of the waveform before (after) the resonance, resulting in a mismatch as large as $\delta \simeq 0.5$. 

The two examples of the fine transition show two qualitatively different behaviors. The example on the left shows a frequency evolution that can be fit well by a vacuum waveform. This is because the gravitational waves that enter the detector are emitted away from the resonance for this particular choice of parameters. The mismatch almost vanishes in this case. The example on the right shows a scenario in which the observed gravitational waves are emitted near the fine resonance. The vacuum waveform is insufficient to fit the frequency evolution, resulting in a large mismatch. 

\begin{figure*}
\centering
\includegraphics[width=0.8\textwidth]{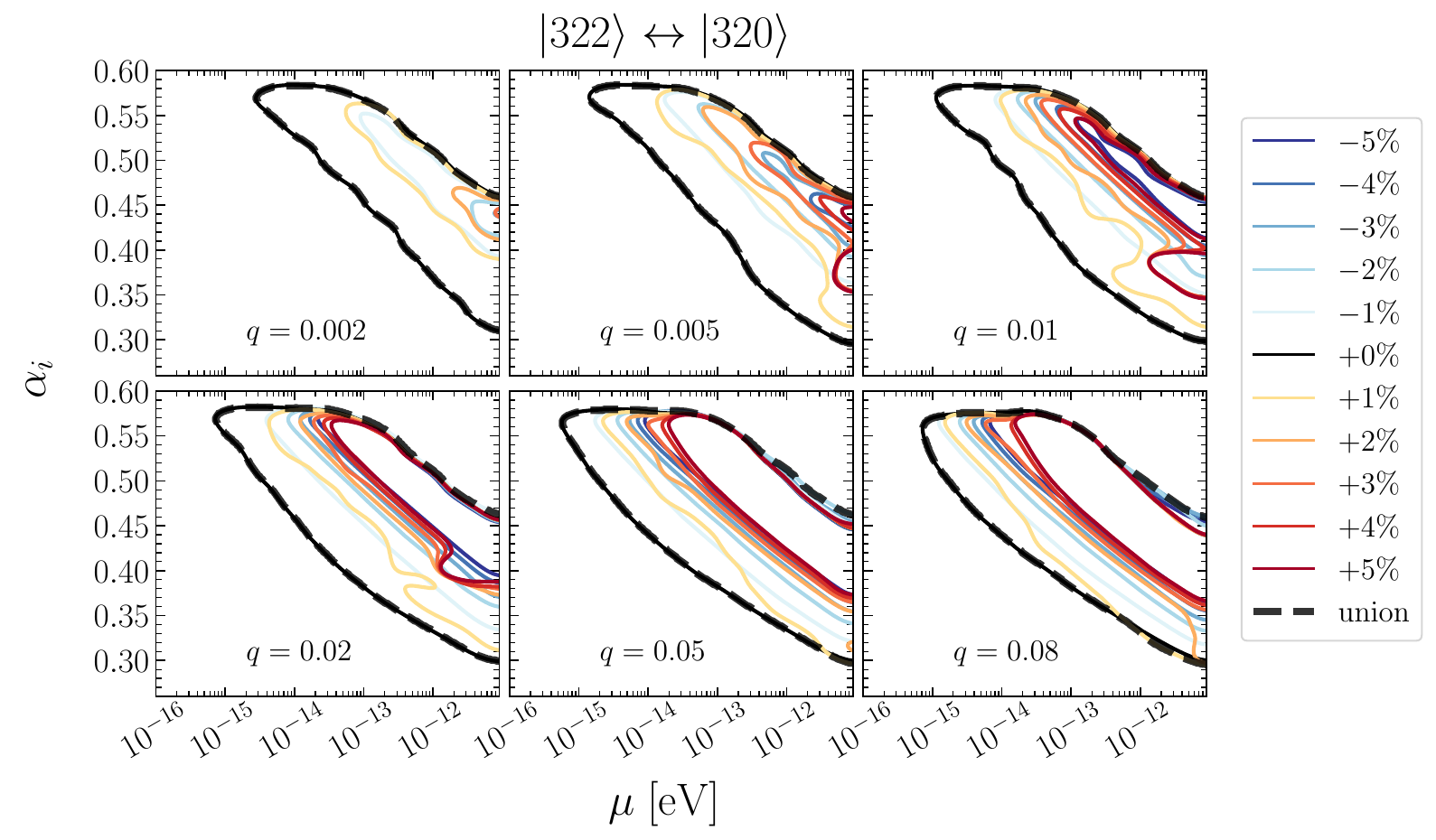}
\includegraphics[width=0.8\textwidth]{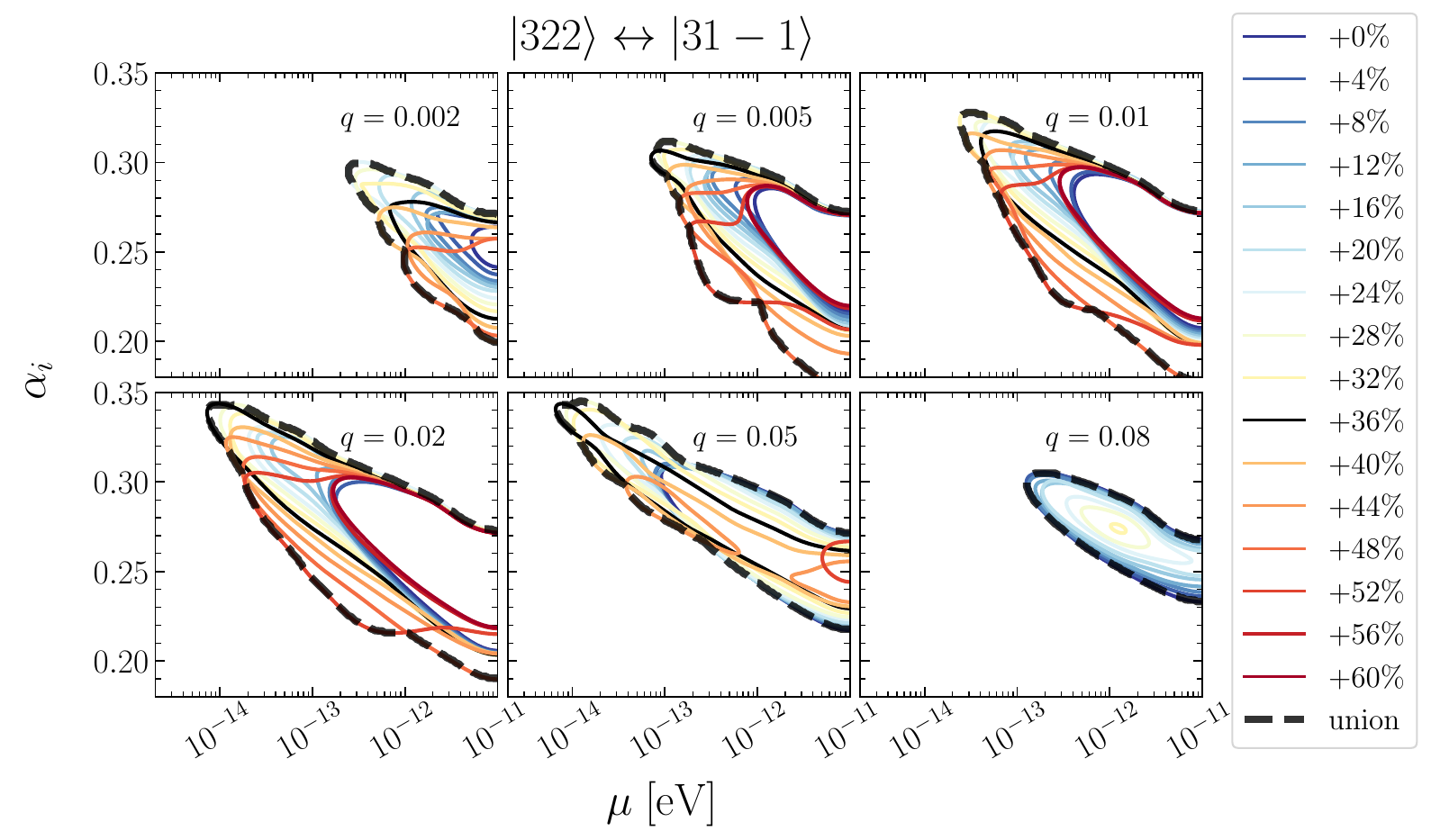}
\caption{The contours of mismatch satisfying the detectability criterion \eqref{detectability_crit} at $2\sigma$ confidence. The fitting factor is computed with different choices of the misalignment frequency factor $c$, defined via $f(t_0) = c (\Omega_0/\pi)$. We label these choices as $100(c-1)\%$. The thick dashed black contours show the union of all the other contours. (Top) the results for the hyperfine resonance. The value of $c$ that contributes to most of the sensitivity is $c\simeq 1$. For chosen values of $q$, LISA might be able to test the mass of ultralight particles down to $\mu  \simeq 10^{-15}\eV$ with the best coverage obtained for large values of $q=0.05$, and $0.08$. (Bottom) the results for the fine resonance. The most relevant value is $c \simeq 1.36$ but the precise value depends on $q$, $\mu$, and $\alpha_i$. In this case LISA could probe the mass of ultralight particles down to $\mu\simeq 10^{-14}\eV$, with best coverage of the parameter space for values of $q$ smaller than in the hyperfine case, i.e. $q=0.01$ and $0.02$, due a larger surviving cloud mass for lower mass ratios.}
\label{fig:mis_scan_qs}
\end{figure*}

As already discussed in the main text, the mismatch depends on whether the observed gravitational waves are emitted near the resonance. In our analysis, this is parameterized by the free parameter $c$ defined by $f(t_0) = c (\Omega_0 / \pi)$. To assess the sensitivity of our results to the precise value of $c$ and also $q$, we compute the mismatch for different values of $(c,q)$ in Figure~\ref{fig:mis_scan_qs}. The contours show the region of the parameter space in which the mismatch satisfies the detectability criterion~\eqref{detectability_crit} at $2\sigma$. Each colored contour assumes a distinct value of $c$. The percentage denotes the deviation of $c$ from $c=1$. The hyperfine transition is sensitive to $c$; a small deviation away from the resonance results in a large reduction in the mismatch. The fine transitions are less sensitive to $c$. This might be attributed to the stronger influence of the $|300\rangle$ state during evolution. 

\section{Relativistic Corrections}\label{app:relativistic}

In the non-relativistic limit, the wave function of the gravitational atom is well approximated by a hydrogenic wave function. However, at large $\alpha$, the wave function receives relativistic corrections. These effects could have an impact on the self-gravity corrections and in general matrix elements that couple different states. We investigate the difference between the hydrogenic wave functions and the ones obtained by solving the equation of motion for the scalar field in Boyer-Lindquist coordinates on the Kerr background. Our discussion closely follows Dolan \cite{Dolan:2007mj}. 

The Klein-Gordon equation is
\begin{align}
    (\square - \mu^2 ) \phi  = 0 \ .
\end{align}
One can decompose the field $\phi$ as 
\begin{align}
   \phi= \frac{1}{\sqrt{2\mu}}e^{-i\omega t}e^{im\phi}S_{\ell m}(\theta)R_{\ell m}(r) + {\rm h.c.}\ .
\end{align}
The equation of motion in spherical coordinates gets decomposed as
\begin{widetext}
\begin{align}
    &0 = \frac{d}{dr}\left( \Delta \frac{dR_{\ell m}}{dr}\right) + \bigg[\frac{\omega^2(r^2+a^2)^2-4mGM \omega a r + m^2 a^2 }{\Delta}
    -(\omega^2a^2 + \mu^2 r^2 +\Lambda_{\ell m})\bigg]R_{\ell m}
\\
    &0 = \frac{1}{\sin\theta}\frac{d}{d\theta}\left( \sin\theta  \frac{dS_{\ell m}}{d\theta}\right) 
    + \bigg[  \kappa^2 \cos^2\theta\ 
    -\frac{m^2}{\sin^2\theta} +\Lambda_{\ell m} \bigg]S_{\ell m}
\end{align}
\end{widetext}
where $\kappa^2 = a^2(\omega^2 -\mu^2)$ is the degree of spheroidicity. The energy and angular eigenvalues $(\omega, \Lambda_{\ell m})$ are unknown. In the limit $a\to 0$,
the spheroidal harmonics reduce to spherical harmonics $Y_{\ell m}$ and $\Lambda _{\ell m} \to \ell (\ell+1)$ to the angular momentum eigenvalues in the hydrogen atom. For values up to $\kappa \sim \ell$, the  expansion $\Lambda_{ \ell m}= \ell (\ell+1) + \sum_{j=1}^{6}  f_j \kappa^{2j}$, with appropriate coefficients $f_j$ tabulated by Seidel \cite{Seidel:1988ue},
is a good approximation. 

With the expected behavior of radial and angular functions at the boundary, we may look for a solution of the following form:
\begin{align}
    R_{\ell m}(r) &\propto  \frac{(r-r_+)^{-i\sigma}} {(r-r_-)^{-i\sigma -\chi +1}} e^{Qr} \sum_{n=0}^{\infty} a_n^{(r)} \left( \frac{r-r_+}{r-r_-}\right)^n, \\
    S_{\ell m}(\theta) &= (1-u)^{|m|/2} (1+u)^{|m|/2} e^{\kappa u} \sum_{n=0}^{\infty}a_n^{(\theta)} (1+u)^n\ .
\end{align}
We define $\sigma = 2r_g(\omega -\Omega_+)r_+ /(r_+-r_-)$, $Q= \sqrt{\mu^2-\omega^2}$ and $\chi = r_g(\mu^2 - 2\omega^2)/Q$. Recall that $\Omega_+ = a_*m /(2 r_+)$ is the angular velocity of the outer horizon $r_+ =r_g(1+\sqrt{1-a_*^2})$. The series coefficients $a_n\equiv a_n^{(r),(\theta)}$ satisfy
\begin{align}
    &\alpha_n a_{n+1} +\beta_n a_n + \gamma_n a_{n-1} = 0 \ ,\\
    &\alpha_0 a_1 +\beta_0 a_0 = 0 
\end{align}
with expressions for $\{\alpha_n,\beta_n,\gamma_n\}$ for both cases given in \cite{Leaver:1985ax, Dolan:2007mj}. One can rewrite these relations as a continued fraction:
\begin{align}
    \beta_0  - \dfrac{\alpha_0 \gamma_1}{\beta_1 - \dfrac{\alpha_1 \gamma_2}{\beta_2 - \dots}} = 0\ .
\end{align}
The above continued fraction, truncated at $n_{\rm max}\sim {\cal O}(10^2 \, \textrm{--} \, 10^3)$, is solved simultaneously for the radial and angular coefficients. The eigenvalues $(\omega,\Lambda_{\ell m})$ are obtained, from which we get the relativistic spectrum of the cloud. 

From the radial and anglar coefficients, we construct the radial function $R_{\ell m}(r)$ as well as angular function $S_{\ell m}(\theta)$. The wave function can be written as $\psi_{n\ell m} (\boldsymbol x) = e^{im\phi}R_{\ell m}(r) S_{\ell m}(\theta)$ with normalization $1 = \int d^3x \,|\psi_{n\ell m}(\boldsymbol x)|^2$. These relativistic wave functions are then used to compute the matrix elements. 

\begin{figure*}
\centering
\includegraphics[width=0.65\textwidth]{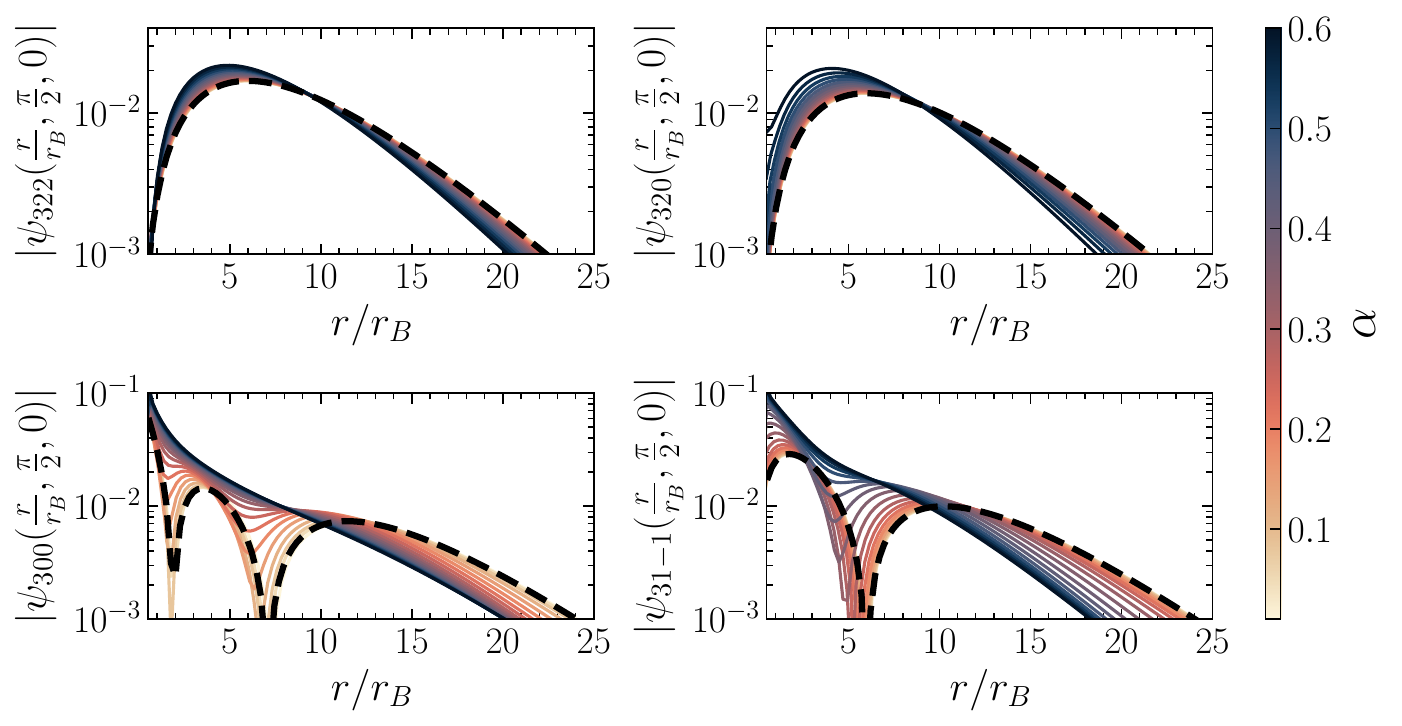}
\caption{Relativistic wave functions evaluated on the equatorial plane at different radii for the $\{|322\rangle,|320\rangle,|300\rangle,|31-1\rangle \}$ states. The colors indicate different values of $\alpha$. The thick dashed line shows the non-relativistic wave functions.}
\label{fig:wave function}
\end{figure*}

\begin{figure*}
\centering
\includegraphics[width=0.65\textwidth]{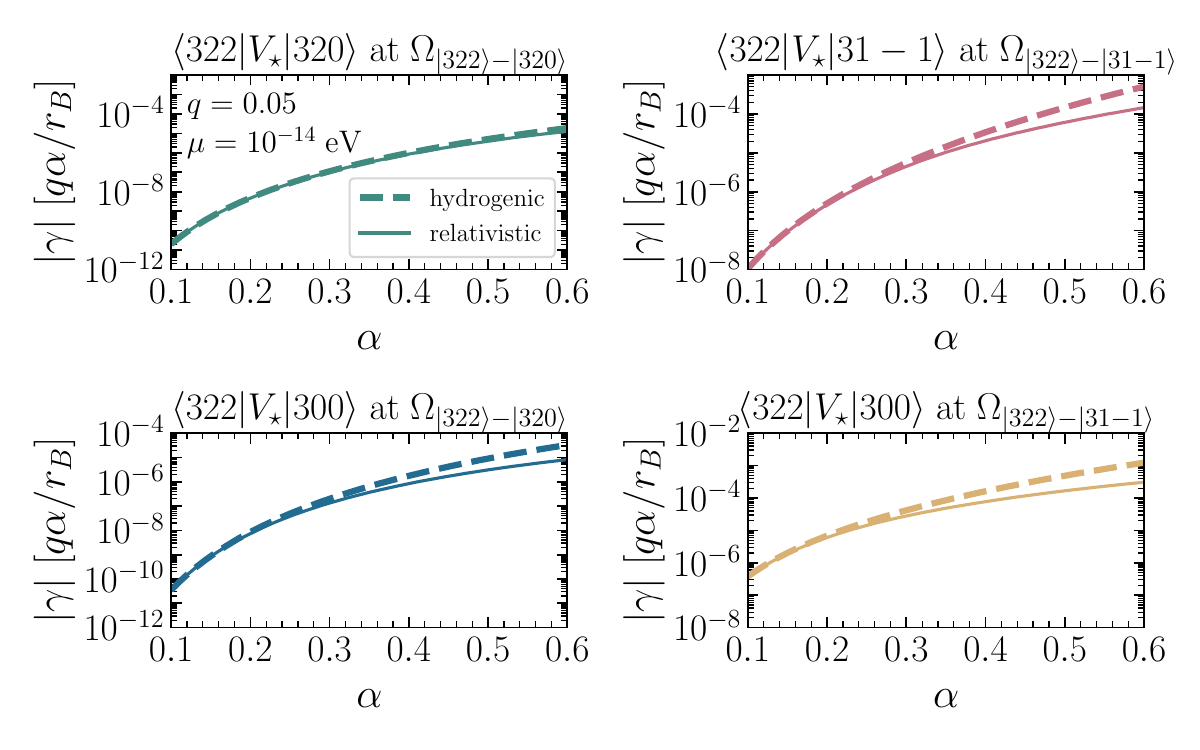}
\caption{Comparison of matrix elements of interest computed at different resonance frequencies in the hydrogenic approximation (thick dashed) and with relativistic wave functions (solid lines).}
\label{fig:relativistic_matrix}
\end{figure*}

We show the relativistic wave functions for the $\{|322\rangle,|320\rangle,|300\rangle,|31-1\rangle \}$ states in Figure~\ref{fig:wave function}. For small values of $\alpha$, the wave functions match the hydrogenic one. For larger $\alpha$, the wave functions become narrower, with the peak shifting closer to the black hole. The relativistic corrections of the wave functions associated with the hyperfine transitions are mild and so are the corrections to the matrix elements. This is confirmed by the explicit calculation shown in Figure~\ref{fig:relativistic_matrix}. 

 
\bibliography{ref}
\bibliographystyle{utphys}
\end{document}